\PassOptionsToPackage{authoryear}{natbib}
\PassOptionsToPackage{disable}{todonotes}
\documentclass[manuscript,screen]{acmart}
\usepackage[colorinlistoftodos,prependcaption,textsize=tiny]{todonotes}
\usepackage{hyperref}
\usepackage{multirow}
\usepackage{caption}
\usepackage{subcaption}

\usepackage{etoolbox}
\usepackage{expl3}
\usepackage{array}
\usepackage{booktabs}
\usepackage{calc}
\usepackage{euscript}
\usepackage{extarrows}
\usepackage[bottom,hang,flushmargin]{footmisc}
\usepackage{graphicx}
\usepackage{graphbox}
\usepackage{listings}
\usepackage{microtype}
\usepackage{multicol}
\usepackage{siunitx}

\usepackage{soul}
\setulcolor{black}
\setstcolor{black}
\sethlcolor{black}

\usepackage{subcaption}
\usepackage{textcomp}
\usepackage[absolute,overlay]{textpos} 
\TPoptions{showboxes=false}
\TPMargin*{2pt} 
\usepackage{upquote}
\usepackage{verbatim}

\usepackage{catchfilebetweentags}
\makeatletter
\patchcmd{\CatchFBT@Fin@l}{\endlinechar\m@ne}{}{}{}
\makeatother

\makeatletter
\@ifclassloaded{acmart}{}{
  \usepackage{amsmath,amssymb,amsthm}
}
\@ifclassloaded{beamer}{} 
{%
  \usepackage[inline,shortlabels]{enumitem}
  \setlist[itemize]{topsep=0pt,itemsep=0pt}

  \usepackage{todonotes}

  \usepackage{hyperref}
  \hypersetup{
    colorlinks=true,
    linkcolor=white,
    urlcolor=blue
    final,                   
    hidelinks,               
    breaklinks=true,         
    anchorcolor=blue,        
  }

  \newcommand{\inlinetodo}[1] {\todo[inline]{#1}\noindent}
  \newcommand{\reference}[1] {\footnotesize{\textbf{Reference:} #1}}
  \newcommand{\setlistspacing}[3] {}

  \newcommand{\showframe}[1] {%
    \begin{center}
      \framebox{\includeslide[width=8cm]{#1}}
    \end{center}
  }

  \newcommand*{\titleslide}[1] {}
}
\makeatother

\definecolor[named]{vlightgrey}{gray}{0.9}
\definecolor[named]{beamergreen}{rgb}{0.2,0.5,0.5}
\definecolor[named]{mygreen}{rgb}{0.4,0.9,0.7}
\definecolor[named]{myred}{rgb}{1,0.7,0.5}

\providecommand{\amber}{}
\providecommand{\blue}{}
\providecommand{\brown}{}
\providecommand{\green}{}
\providecommand{\grey}{}
\providecommand{\red}{}
\providecommand{\white}{}

\renewcommand{\amber}[1] {\textcolor[rgb]{1.00,0.50,0.00}{#1}}
\renewcommand{\blue}[1] {\textcolor{blue}{#1}}
\renewcommand{\brown}[1] {\textcolor{Mahogany}{#1}}
\renewcommand{\green}[1] {\textcolor[rgb]{0,0.7,0}{#1}}
\renewcommand{\grey}[1] {\textcolor{gray}{#1}}
\newcommand{\mygreen}[1] {\textcolor{mygreen}{#1}}
\newcommand{\myred}[1] {\textcolor{myred}{#1}}
\renewcommand{\red}[1] {\textcolor{red}{#1}}
\renewcommand{\white}[1] {\textcolor{white}{#1}}

\ifdefined\labelenumiii
  \renewcommand{\labelenumiii}{(\roman{enumiii})}
\else
  \newcommand{\labelenumiii}{(\roman{enumiii})}
\fi

\newfont{\helvetica}{phvr8t at 10pt}
\newfont{\helveticasmall}{phvr8t at 9pt}
\newfont{\helveticabig}{phvr8t at 22pt}
\newfont{\helveticao}{phvro8t at 11pt}

\newcommand{\smallsc}[1] {{\footnotesize\textsc{#1}}}
\newcommand{\setfontsizeandspacing}[2] {\fontsize{#1}{#2}\selectfont}

\makeatletter
\def\lst@visiblespace{\mbox{\kern.1em\vrule height.6ex}%
  \vbox{\hrule width1ex}%
  \hbox{\vrule height.6ex}%
}
\makeatother
\newcommand{\lstsetforpython}
{\lstset{language=Python,frame=single,frameround=tttt,framesep=3pt,backgroundcolor=\color{mygreen!20},rulecolor=\color{mygreen}}}

\usepackage{tikz} 
\usetikzlibrary{arrows.meta}
\usetikzlibrary{calc}
\usetikzlibrary{decorations.text}
\usetikzlibrary{positioning}
\usetikzlibrary{shapes, shapes.callouts}
\usetikzlibrary{trees}
\tikzset{
  >={Stealth}, 
  invisible/.style={opacity=0},
  visible on/.style={alt=#1{}{invisible}},
  alt/.code args={<#1>#2#3}{%
    \alt<#1>{\pgfkeysalso{#2}}{\pgfkeysalso{#3}} 
  },
}

\newcolumntype{C}[1]{>{\centering\arraybackslash}m{#1}}   
\newcolumntype{L}[1]{>{\raggedright\arraybackslash}m{#1}} 
\newcolumntype{R}[1]{>{\raggedleft\arraybackslash}m{#1}}  

\newcolumntype{+}{>{\global\let\currentrowstyle\relax}}
\newcolumntype{^}{>{\currentrowstyle}}
\newcommand{\rowstyle}[1]{\gdef\currentrowstyle{#1}#1\ignorespaces}

\makeatletter
\@ifundefined{htlatex}{%
  \usepackage{ctable}
  \setupctable{%
    botcap, 
    captionskip=2pt, 
    mincapwidth=2in, 
    framebg=1 1 1, 
    framefg=0 0 0, 
    framerule=0pt, 
    framesep=0pt,  
  }
}
\makeatother

\usepackage{colortbl}

\def\citea{\citet}

\newcommand{\parab}[1] {\bigskip \noindent \textbf{#1}~}
\newcommand{\param}[1] {\medskip \noindent \textbf{#1}~}
\newcommand{\paras}[1] {\smallskip \noindent \textbf{#1}~}

\newcommand{\setspacing}[1] {\linespread{#1}\selectfont}
\def\spacing {\setspacing}

\newlength{\parindentincr}
\newcommand{\addindent} {\addtolength{\parindent}{5mm}}
\newcommand{\indentby}[1] {
  \settowidth{\parindentincr}{#1}
  \addtolength{\parindent}{\parindentincr}}
\newcommand{\deindent} {\addtolength{\parindent}{-5mm}}
\newcommand{\deindentby}[1] {
  \settowidth{\parindentincr}{#1}
  \addtolength{\parindent}{-\parindentincr}}

\newcommand{\blank}[1] {\rule{#1}{0.5pt}}

\newcounter{zzlinei}
\newcommand{\blanklines}[1] {%
  \setcounter{zzlinei}{0}
  \whileboolexpr%
    {test{\ifnumcomp{\value{zzlinei}}{<}{#1}}}
    {\vspace{8mm}\hrulefill\par \addtocounter{zzlinei}{1}}
  }

\newcommand{\cl} {\centerline}
\newcommand{\command}[1] {\textcolor{Mahogany}{\texttt{#1}}}
\newcommand{\dash} {\blank}
\newcommand{\definedas} {\ensuremath{\triangleq\ }}
\newcommand{\dfslab} {\url{http://www.isical.ac.in/~dfslab}}
\newcommand{\emptybox} {\raisebox{3.3pt}{$\boxed{}$}}
\newcommand{\checkedbox} {\raisebox{3.3pt}{$\boxed{}$}\hspace{-7pt}$\checkmark$}
\newcommand{\greentick} {\includegraphics[scale=0.1]{green-tick}}
\providecommand{\half}{}
\renewcommand{\half} {\ensuremath{\frac{1}{2}}}
\renewcommand{\hl}[2][YellowOrange!50] {\colorbox{#1}{#2}}
\newcommand{\hlg}[1] {\colorbox{mygreen}{#1}}
\newcommand{\hlr}[1] {\colorbox{myred}{#1}}
\newcommand{\important}[1] {\centerline{\framebox{\textcolor{red}{#1}}}}
\newcommand{\intersection} {\cap}
\newcommand{\latex} {\LaTeX\ }
\newcommand{\lb} {\linebreak}
\renewcommand{\marks}[1] {\hspace*{\fill}{[#1]}}
\newcommand{\mc} {\multicolumn}
\newcommand{\mybox}[1] {\centerline{\framebox{#1}}}
\newcommand{\mycross} {\includegraphics[scale=0.1]{red-cross}}
\newcommand{\myellipsis}[3] {\newcount \secondindex
  \secondindex #2\relax
  \advance\secondindex +1\relax
  \ensuremath{{#1}_{#2},}
  \ensuremath{{#1}_{\the\secondindex},}
  \ensuremath{\ldots,}
  \ensuremath{{#1}_{#3}}%
}
\newcommand{\mysignature} {\includegraphics[scale=0.2]{signature}}
\newcommand{\mystrut}[1] {\rule[- #1 * \real{0.3}]{0pt}{#1}}
\newcommand{\otoprule} {\midrule[\heavyrulewidth]}
\providecommand{\overbar}{}
\renewcommand{\overbar} {\overline}
\newcommand{\partialinput}[2] {\ExecuteMetaData[#2]{#1}}
\NewCommandCopy{\pic}{\includegraphics}
\newcommand{\pipe} {~\ensuremath{\mid}~}
\newcommand{\pto} {\vfill\hspace*{\fill}\textsc{p.t.o.} \newpage}
\newcommand{\ra} {\ensuremath{\rightarrow}}
\newcommand{\Ra} {\ensuremath{\Rightarrow}}
\newcommand{\rcomment}[1] {\bigskip\color{blue}#1\color{black}}
\newcommand{\response} {\medskip\textbf{Response:~}}
\newcommand*{\rectgolla}[2][red]{
  \setlength{\fboxrule}{1pt}
  \fcolorbox{#1}{white}{#2}
} 
\newcommand{\sbcb} {\;\!} 
\renewcommand{\show} {\visible}
\renewcommand{\so} {\st}
\newcommand{\strikethrough} {\st}
\newcommand{\tb} {\textbackslash}
\newcommand{\torf} {\hfill \textsc{true / false}}
\newcommand{\tpgridon}{\TPShowGrid*{15}{12}} 
\newcommand{\ttilde} {\textasciitilde}
\newcommand{\undertilde}[1] {\mbox{\Large \raisebox{-5mm}{$\displaystyle\stackrel{#1}{\text{\textasciitilde}}$}}}
\newcommand{\union} {\cup}
\NewCommandCopy{\oldvec}{\vec}
\renewcommand{\vec}[1] {\oldvec{\vphantom{b}#1}}
\newcommand{\xor} {\ensuremath{\oplus}}

\newcommand{\anchorpoint}[1]{\tikz[overlay,remember picture] \node (#1) {};}

\newcommand*{\floatingnote}[5][1-]{
  \begin{tikzpicture}[overlay,remember picture]
    \pgftransformshift{\pgfpointanchor{current page}{center}}
    \path<#1> #2
    node[rectangle,rounded corners=4pt,fill=#4,opacity=.5]
    {\parbox{#3}{\raggedright #5}}
    ;
  \end{tikzpicture}
}

\newcommand*{\golla}[4][ellipse]{
  \begin{tikzpicture}[overlay,thick]
    \node () at #2 [shape=#1,draw=red,text height=#4] {~\hspace{#3}~} ;
  \end{tikzpicture}
}
 
\newcommand*{\namedgolla}[5][red]{
  \begin{tikzpicture}[overlay,remember picture,thick]
    \node (#5) at #2 [shape=ellipse,draw=#1,text height=#4] {~\hspace{#3}~} ;
  \end{tikzpicture}
}

\newcommand*{\popupnote}[4][1-]{
  \begin{tikzpicture}[overlay,remember picture]
    \pgftransformshift{\pgfpointanchor{current page}{center}}
    \path<#1> (#2) ++#3
    node[anchor=west,rectangle callout,rounded corners=4pt,fill=blue!50,
         opacity=.5,callout absolute pointer={(#2)}] {#4};
  \end{tikzpicture}
}

\newcommand*{\popupnotewithap}[5][1-]{
  \begin{tikzpicture}[overlay,remember picture]
    \pgftransformshift{\pgfpointanchor{current page}{center}}
    \path<#1> (#2) ++#3
    node[anchor=#5,rectangle callout,rounded corners=4pt,fill=blue!50,
         opacity=.5,callout absolute pointer={(#2)}] {#4};
  \end{tikzpicture}
}

\makeatletter
\newread\pin@file
\newcounter{pinlineno}
\newcommand\pin@accu{}
\newcommand\pin@ext{pintmp}
\newcommand*\dummypartialinput [3] {%
  \IfFileExists{#3}{%
    \openin\pin@file #3
    \setcounter{pinlineno}{1}
    \@whilenum\value{pinlineno}<#1 \do{%
      \read\pin@file to\pin@line
      \stepcounter{pinlineno}%
    }
    \addtocounter{pinlineno}{-1}
    \let\pin@accu\empty
    \begingroup
    \endlinechar\newlinechar
    \@whilenum\value{pinlineno}<#2 \do{%
      \readline\pin@file to\pin@line
      \edef\pin@accu{\pin@accu\pin@line}%
      \stepcounter{pinlineno}%
    }
    \closein\pin@file
    \expandafter\endgroup
    \scantokens\expandafter{\pin@accu}%
  }{%
    \errmessage{File `#3' doesn't exist!}%
  }%
}
\makeatother

\renewcommand{\citet}[2][] {\citeauthor{#2} (\citeyear{#2})~\cite[#1]{#2}}

\AtBeginDocument{%
  \providecommand\BibTeX{{%
    Bib\TeX}}}

\setcopyright{acmlicensed}
\copyrightyear{2018}
\acmYear{2018}
\acmDOI{XXXXXXX.XXXXXXX}
\acmConference[Conference acronym 'XX]{Make sure to enter the correct
  conference title from your rights confirmation email}{June 03--05,
  2018}{Woodstock, NY}
\acmISBN{978-1-4503-XXXX-X/2018/06}

\begin{document}
\title{Explainability of Text Processing and Retrieval Methods: A Survey}

\author{Sourav Saha}
\affiliation{%
	\institution{Indian Statistical Institute}
	\city{Kolkata}
	\country{India}
}
\email{sourav.saha\_r@isical.ac.in}

\author{Debapriyo Majumdar}
\affiliation{%
	\institution{Indian Statistical Institute}
	\city{Kolkata}
	\country{India}
}
\email{debapriyo@isical.ac.in}

\author{Mandar Mitra}
\affiliation{%
	\institution{Indian Statistical Institute}
	\city{Kolkata}
	\country{India}
}
\email{mandar@isical.ac.in}

\begin{abstract}
  Deep Learning and Machine Learning based models have become extremely
  popular in text processing and information retrieval,
  owing to their remarkable effectiveness. However, the
  complex non-linear structures underlying these models
  make them largely inscrutable. A significant body of research has focused
  on increasing the transparency of these models. This article provides a
  broad overview of research on the explainability and interpretability of
  information retrieval methods, focusing primarily on both
    neural models for document ranking, and retrieval-augmented generation
    systems. A significant part of this research is inspired by, and
    strongly related to, similar work done in the area of natural language
    processing (NLP). For completeness, therefore, we also briefly review
    (in the Appendices) some recent studies from the NLP community on
    explaining word embeddings, sequence modeling, attention modules,
  transformers, and BERT. The concluding section suggests some possible
  directions for future research on this topic.
\end{abstract}



\maketitle


\section{Introduction}
\label{sec:intro}
Given the volume of information that we routinely encounter, we have become
acutely dependent on technology that processes the available
information, classifies it on the basis of diverse criteria, selects useful
information from this barrage, and presents it in a form that is easy to
consume and assimilate. It is not enough for such technology to be merely effective: as humans, we would also like to understand why our
tools behave the way they do. Naturally, a substantial body of research has
focused on methods for providing high-level, human-understandable
\emph{explanations} or \emph{interpretations} of the inner workings of, as
well as the results produced by, information processing systems.

Up to about 12 years ago, commonly used information processing tools
generally extracted a \emph{feature vector} --- a list of features along
with their associated numerical weights --- from the form in which the
information was initially provided: structured or unstructured text,
images, audio, video, etc. These features typically corresponded to
human-understandable properties or attributes of the original information.
We confine our discussion in this article to textual information only. For
text, features would include words and phrases occurring in the text, their
counts, morphology, parts of speech, and various grammatical relationships. 
These features were chosen on the basis of
linguistic considerations; they were, in a sense, interpretable by design.
A number of studies (some of which are discussed in more detail in
Section~\ref{sec:document-ranking}) did focus on understanding feature
weights, their behaviour, and the mathematical~/ statistical models
underlying their computation and use, but they were not explicitly
designated as \emph{explainability}~/ \emph{intepretability} studies.

Over the past few years, advances in Deep Machine Learning (ML) techniques
--- in particular, the potent combination of Large Language
  Models (LLMs) and Retrieval Augmented Generation (RAG) systems ---  have
revolutionised the state of the art in textual information processing (as
well as a number of other areas).
  The traditional paradigm of \emph{document ranking} ---
  represented by ``ten blue links'' --- is increasingly being replaced by
  interactive and conversational forms of information
  access~\cite{fntir-conv-ir}.
These radical improvements have come at a cost: text is no longer simply
represented in terms of easily interpretable features. Instead, early in
the information processing pipeline, text is converted to so-called
``dense'' vectors in $\mathbb{R}^n$; the dimensions of this space are no longer
easily understood. This representation is then processed by ML systems that
have large and complex architectures, involving many parameters, often
numbering in the billions. As a fallout, there has been a marked
increase 
in the attention devoted to providing high-level, intuitive explanations of
these models and methods.


Thus, this seems to be an appropriate time to present an organised summary
of the major research efforts in the area of explainability in textual
information processing and retrieval. Indeed, several such general surveys
have already been
published~\cite{sogaard2021explainable,posthoc-interpret-neural-nlp,10.1145/3529755,explain-ir-avishek-anand-survey},
in addition to more specialised surveys, such
as~\cite{belinkov:2019:tacl,position-info-transformer}. Our objective is to
complement these existing surveys by describing more recent work in areas
already covered there, and by focussing on areas not addressed by them.
Before discussing the relation between this survey and the existing ones
in Section~\ref{sec:related-work}, we first briefly describe how we collected the set of papers
reviewed in this survey.

\subsection*{{Scope of this survey}}
\label{sec:scope}
Text processing and retrieval problems are studied by researchers from the
Information Retrieval (IR) and Natural Language Processing (NLP)
communities, as well as from the broader Machine Learning / Deep Learning
(DL) communities. Thus, to ensure comprehensive coverage of explainability
and interpretability studies relating to IR and NLP, we identified a wide
range of highly-regarded venues including \emph{SIGIR}, \emph{CIKM},
\emph{WSDM}, \emph{WWW}, the \emph{ACL} family of conferences and
journals, 
\emph{AAAI}, \emph{ICML}, \emph{ICLR}, \emph{NeurIPS}, \emph{FAccT}, ACM
\emph{TOIS} and \emph{IPM} (Information
Processing \& Management).

The principal terms used in our search were variants of the keywords
\emph{explanation} (including \emph{explain*}) and
\emph{interpretability} (viz., \emph{interpret*}). These terms have been
variously defined within the community. For example, according to \citet{lime},
explaining a model may be defined as presenting textual or visual cues that
provide qualitative understanding of the relationship between the
components of an input instance (e.g. words in text), and the model's output
for that instance. \citet{doshivelez2017towards} define interpretability
more generally as simply ``the ability to explain or to present in
understandable terms to a human''. We adopt the broad sense conveyed by
these definitions, and use the terms explainability and interpretability
interchangeably in this survey.

Our preliminary search revealed a substantial body of work on explainability in Recommender Systems and Computer Vision. However, we excluded papers from these areas unless they directly contributed to follow-up work in IR or NLP. While IR systems and Recommender Systems share similarities in their goal of ranking items to meet users' information needs, often with or without explicit queries, this survey focuses specifically on the retrieval of \emph{text} documents. As text-focused IR methods are deeply intertwined with NLP techniques, we concentrate on the explainability of IR and NLP to maintain this focus.
We include papers published from 2004 to 2025 in this set.
Some influential papers published long before 2004 were also added.
From the longlisted set of
papers, a subset of 60 papers was selected for more detailed discussion in this article.
We also provide short overviews, insights of several additional works; however, these are not included in the reported count.


\paras{Organisation of the survey:} The remainder of this survey is
organised as follows. Section~\ref{sec:related-work} lists existing surveys
that cover explainability in textual IR and NLP, and describes the relation
between them and the current article. Section~\ref{sec:basic-terminology}
introduces the basic terminology used in the literature on explainability,
and presents a schema for classifying explainability research. The studies
summarised in the rest of the article are broadly organised using this
schema. Section~\ref{baseline-methods} reviews some well-known approaches
to explainability that are most commonly used as baselines in recent work.
  Section~\ref{sec:eval-expl} briefly describes the metrics
  that are currently commonly used to measure the quality of explanations.
  Even though metrics are 
  crucial to measuring progress, it is worth noting that evaluation
  protocols for explainability are yet to achieve a satisfactorily mature
  status.
Section~\ref{sec:document-ranking} is devoted to
a detailed discussion of explainable document / passage ranking, and is
itself divided into subsections by topic. Within each of these sections and
subsections, different studies have been discussed in chronological order.
%
 The expanding field of explainability for Retrieval
  Augmented Generation (RAG) Systems is discussed in a separate section
  (Section~\ref{sec:explain-rag}).
We conclude in Section~\ref{sec:future} with a few research questions that
we believe merit further investigation.

For completeness, we also provide a set of appendices (\ref{sec:embeddings}
through \ref{sec:transformers-bert}) that are intended as supplements to
existing surveys (specifically those by
\citet{posthoc-interpret-neural-nlp} and \citet{10.1145/3529755}). These
appendices cover recent work (not discussed in
\cite{posthoc-interpret-neural-nlp,10.1145/3529755}) on explainability for
individual elements of deep text processing pipelines, viz., word
embeddings, sequence models, attention mechanisms, and transformer
architectures.
%

\section{Related Surveys}
\label{sec:related-work}
In a monograph, ~\citet{jimmy-lin-book} 
provide a comprehensive survey of neural IR models, covering topics from
multi-stage ranking architectures to dense retrieval techniques. The
authors observe that IR models may be stratified into three fairly
well-separated levels on the basis of effectiveness: transformer-based
ranking strategies represent a definite advance over pre-transformer Neural
Ranking Models (NRMs), which in turn improve significantly over
``traditional''\footnote{In this survey, we use this adjective to refer to
  all non-neural IR models, including relatively recently proposed ones.} statistical ranking models. An intuitive (but precise)
grasp of what leads to these quantum jumps in effectiveness seems eminently
desirable.
%
This, along with a number of other factors, has led to much interest in the
explainability of NRMs. There has always been a substantial
overlap between the IR and NLP communities. With the recent success of
neural methods, the boundary between the disciplines has been blurred even
further. Not surprisingly, therefore, research in explainable IR (ExIR) is
significantly influenced by similar research in NLP, which in turn draws
heavily from explainability studies in ML in general.

Our search for a comprehensive summary of research on this broad subject
led us to two articles that provide an overview of attempts to
explain neural models for
NLP~\citep{posthoc-interpret-neural-nlp,10.1145/3529755}. While
\citet{posthoc-interpret-neural-nlp} focus on \emph{post-hoc}\footnote{In
  Section~\ref{sec:basic-terminology}, we provide a review of this and
  other basic terms commonly used in the context of interpretability /
  explainability.} methods for interpreting neural NLP methods, the main
emphasis in (\citet{10.1145/3529755}) is on the explainability of word
embedding vectors, recurrent neural networks (RNNs) and attention
mechanisms.

This being an area of active research, a non-trivial body of work exists
that has not been covered by the above surveys, either because the work was
published later, or because its topic was outside the specific scope of the
surveys. In particular, until very recently, there were no similar
overviews addressing ExIR in detail. The current study started as an
attempt to address that lacuna. 
\citet{explain-ir-avishek-anand-survey} have published an article focused
on this specific topic. There is, inevitably, a good deal of overlap
between \citeauthor{explain-ir-avishek-anand-survey}'s survey and
Section~\ref{sec:document-ranking} of this article, 
but we also cover a fair amount of work that is not discussed by
them. Further, since our
survey was prepared independently, we believe that it provides a somewhat different
perspective on ExIR.
Finally, given that ExIR often draws inspiration from explainability
for NLP, we review (in Sections~\ref{sec:embeddings} through
\ref{sec:transformers-bert}) recent work on Explainable NLP that was not
covered by the earlier
surveys~\citep{posthoc-interpret-neural-nlp,10.1145/3529755}.
In the rest of this section, we briefly discuss other
related surveys that are either older, or have a scope that does not
substantially overlap with ours.


To the best of our knowledge, \citet{doshivelez2017towards} were the first
to attempt to formalise various notions related to interpretability in ML.
\citet{lipton-16} provides an exposition of the broader aspects of
interpretability in ML and Artificial Intelligence (AI), and why it is
challenging, especially in the modern scenario.
\citet{sogaard2021explainable} examines extant taxonomies of
explainability methods, and carefully constructs a framework for organising
them. The next section draws from these articles to build a glossary of
common terms used in this area.

\citet{belinkov:2019:tacl} cover explainable NLP, and summarise studies
that analyse neural network nodes, and activations across different layers
of deep neural networks, in order to understand what linguistic features are
encoded by neural NLP models. As
mentioned above, many of the discussed methods were based on general
interpretability approaches proposed by the ML community, often within the
context of Computer Vision / Image Processing problems. Signalling the
growing interest in this area, 
\citet{danilevsky-etal-2020-survey} also provide an early overview of
explainability for NLP, describing different explanation categories and
approaches, as well as several visualization techniques.
\citet{10.1162/tacl_a_00440} focus on explanation as a tool for
\emph{debugging} NLP models, with humans in the loop.

As Transformer and BERT-based techniques were widely adopted in the IR / NLP
community, several researchers focused on various aspects of BERT, from its
knowledge-representation power to fine-tuning, pre-training, and
over-parametrisation issues. \citet{rogers-etal-2020-primer} were the first
to survey research in this category, and present a summary of ``what we
know about how {BERT} works.'' The surveys discussed at the beginning of
this section provide a more up-to-date summary of research in this broad area.

There are other broader efforts in Explainable AI (XAI), such as XAIR \cite{Clement2023XAIRAS}, which systematically reviews XAI methods aligned with the software development process. However, this meta-review targets general AI explainability rather than the specific challenges of textual and retrieval systems addressed in our survey.


\section{Categorising Explanation Methods}
\label{sec:basic-terminology}

Methods for explaining IR / NLP models and techniques can be categorised on
the basis of a number of criteria. While introducing these criteria (and
also in the rest of this article), we use the symbols listed in
Table~\ref{tab:notation}.
\begin{table}[h]
  \centering
  \begin{tabular}{l L{0.85\textwidth}}\toprule
    $\mathcal{Q}$ & set $\{ \myellipsis{Q}{1}{m} \}$ of queries \\
    $\mathcal{D}$ & corpus $\{ \myellipsis{D}{1}{n} \}$ of documents \\
    $Q$ & representative of any query from $\mathcal{Q}$, comprising a set
          $\myellipsis{q}{1}{t}$ of terms \\
    $D$ & representative of any document from $\mathcal{D}$ consisting of the terms
          $\myellipsis{d}{1}{l}$ \\
    $\mathcal{C}$ & class of complex models that need explanations \\
    $\mathcal{M}_C$ & representative of any member of $\mathcal{C}$ \\
    $\mathcal{S}$ & class of simple models relatively easily
          understood by humans \\
    $\mathcal{M}_S$ & representative of any instance of $\mathcal{S}$ \\
    $I$ & representative of any instance provided as input to a model \\
    $I_C$ & actual representation of $I$ (the ``feature vector'', often hard 
            to interpret) that is taken as input by $\mathcal{M}_C$ \\ 
    $I_S$ & simple representation of $I$, easily understood by humans,
    often provided as input to $\mathcal{M}_S$ \\\bottomrule
  \end{tabular}
  \caption{Summary of notation used in this article.}
  \label{tab:notation}
\end{table}
In subsequent sections, we try to classify and label the reviewed
approaches using the criteria described below.

\subsection{Model-agnostic vs.\ model-aware explanations}
\label{sec:n-or-a}
Explanations that use information about the internal structure of a model
$\mathcal{M}_C$ are referred to as \emph{model-specific} or \emph{model-aware}. In
contrast, \emph{model-agnostic} explanations treat the explained model as a
black box, and make no assumptions about its internal architecture or
parameters. Model-agnostic methods often submit controlled perturbations of
an input instance $I$ to $\mathcal{M}_C$, and try to gather insights about $\mathcal{M}_C$
based on how its output is affected by these perturbations.

\subsection{Inherent explainability vs.\ post-hoc interpretability}
\label{sec:i-or-p}
Some ML models, e.g., rule-based systems, decision trees and sparse linear
models, are regarded as \emph{inherently explainable}, because it is
relatively simple for a human to trace and describe their inner workings.
Traditional IR models like BM25 are also generally regarded as
\emph{inherently explainable}. Such models comprise the class $\mathcal{S}$ (see
Table~\ref{tab:notation}). Explanations of such models are, naturally,
model-aware. The composition of $\mathcal{S}$ is open to debate, however.
\citet{pmlr-v119-moshkovitz20a} argue that even apparently
well-understood methods with a single parameter, viz., $k$-means and
$k$-medians clustering techniques, may be regarded as ``opaque'', because
these methods define cluster boundaries on the basis of linear combinations
of features, rather than a single feature value. More pertinent to our
context is the debate on whether the attention mechanism can be construed
as providing explanations. Section~\ref{debate-on-attention} discusses this
debate in more detail.

In contrast, \emph{post-hoc explanations} primarily seek to elucidate the
\emph{output}, rather than the inner workings of models.
Visualisations of word embeddings (e.g., using
t-SNE~\cite{van2008visualizing}) are a form of post-hoc ``explanation'':
they do not explain why certain words are mapped to certain vectors,
but once the vectors are generated (i.e., \emph{after the fact}), they
provide insights about the correspondence between proximity in the
semantic and vector spaces. Explanations by example (e.g., ``Model
$M$ predicts class $Y$ for input instance $X$ because $X$ is `similar' to
training instances $\myellipsis{X}{1}{k}$ which were labelled $Y$'') also
fall within this category. Post-hoc explanations may be either model-aware
or model-agnostic.

\subsection{Scope of an explanation: global vs.\ local} 
\label{sec:g-or-l}
\emph{Global} methods explain the overall behaviour of a model.
Global explanations should apply to all (or most) input-output
pairs for a given model. An example of a global explanation in the IR
context is: ``ranking based on query likelihood tends to be biased in
favour of short documents when Jelinek-Mercer smoothing is used instead of
Dirichlet smoothing''~\cite{smucker2005investigation}. On the other hand, a
\emph{local} explanation only seeks to explain a model's decision for a
given input instance.

\subsubsection*{Pairwise vs.\ pointwise vs.\ listwise local explanations}
\label{sec:ppl}
For ranking models, local explanations can in turn be classified into three
sub-categories.
\emph{Pointwise explanations} explain the specific decision made by a model
for a particular input instance. In the IR context, a pointwise explanation
addresses the question of why a document $D$ receives a particular
similarity score for a query $Q$. Apart from ranking models, pointwise
explanations may also be constructed for classifiers.
\emph{Pairwise explanations} are typically offered for systems that rank
items (e.g., documents) based on their relationship to another item (e.g.,
a query). In the IR context, for a given query $Q$ and a document pair
$\left< D_i, D_j \right>$, a pairwise explanation tries to answer why a
model ranks $D_i$ above / below $D_j$ for $Q$.
\emph{Listwise explanations} attempt to provide a \emph{single} explanation
for a complex model $\mathcal{M}_C$'s ranking of the top-$k$ items (documents) related
to a given item (query). We discuss one possible form of such an
explanation in the next section.

\subsection{Nature or form of post-hoc explanations}
\label{sec:nature-or-form}
At their simplest, explanations are observations or statements about how
models behave. Some global explanations are of this form: see
Section~\ref{sec:g-or-l} for an example; more examples (about what
knowledge BERT-based models possess) will be discussed in
Section~\ref{bert-ranking}. Naturally, simple explanations can rarely be
provided for complex neural models. In practice, therefore, the
explanations constructed by researchers appear in diverse forms. Below, we
briefly describe some forms in which explanations are commonly provided.

\subsubsection{Explainable surrogate models}
\label{sec:surrogates}
One widely adopted approach involves trying to construct an $\mathcal{M}_S \in
\mathcal{S}$ that mimics the behaviour of a given complex model $\mathcal{M}_C$. In principle,
$\mathcal{M}_S$, the explainable surrogate, may emulate $\mathcal{M}_C$ on most inputs; it may
then be regarded as a \emph{global} explanation. In practice, it is more
feasible to provide a \emph{local} explanation for a given $I$, by
constructing an $\mathcal{M}_S \in \mathcal{S}$ along with a simple representation $I_S$ of $I$,
such that the behaviour of $\mathcal{M}_C$ in the vicinity of $I$ is 
similar to the behaviour of $\mathcal{M}_S$ on $I_S$.
%
As an example, suppose a neural ranker $\mathcal{M}_C$ generates a ranked list of
documents for a given query $Q$. Then, a listwise explanation for this
output might consist of a traditional document-ranking algorithm, e.g.,
BM25, and a list $L=\{\myellipsis{t}{1}{k}\}$ of \emph{explanation terms}.
The pair $\left\langle \mathrm{BM25}, L \right\rangle$ is interpreted as an explanation
in the following sense: given $Q$, $\mathcal{M}_C$ must be implicitly detecting
matches between documents and the terms in $L$ in the same way that BM25
would, if it were explicitly given $L$ instead of $Q$.

\subsubsection{Feature attribution and rationales}
\label{sec:feat-attr}
Instead of explaining $\mathcal{M}_C$'s behaviour in terms of $\mathcal{M}_S$, researchers have
also attempted to understand $\mathcal{M}_C$ by identifying which features of an
input instance $I$ (i.e., which components of $I_S$ or $I_C$) most strongly
affect $\mathcal{M}_C$'s output for $I$. This approach is called \emph{feature
  attribution}. Several techniques for quantifying and visualising the
contribution of a particular feature towards the prediction made by a model
have been proposed in the literature. These local explanations may be
provided graphically in the form of a heatmap or saliency
map~\cite{saliency-map-simonyanVZ13}, with different colours or shades
representing different degrees of importance of features, or simply as a
list of the most important features along with their importance scores.

A related approach involves the construction of a \emph{rationale}, a
concise representation (or the extracted essence) of $I$ that is (i)~both
short and simple enough to be readily understandable by a human, but
(ii)~for which $\mathcal{M}_C$'s prediction is the same as that for $I$. For example,
a short excerpt from an input text that is sufficient to correctly predict
the sentiment of a sentence comprises a rationale for a sentiment
classification model~\citep{deyoung-etal-2020-eraser}.

\subsubsection{Contrastive explanations}
\label{sec:contrastive}
Contrastive methods~(\citet{jacovi-etal-2021-contrastive})
  aim to elucidate a model's prediction for a particular instance $I$, by
  contrasting it with an alternative, closely related instance $I'$ for
  which the prediction is different. These methods are, by design,
  local in nature. In practice, researchers (e.g.,
  \citet{counterfactual-himabindu}) frequently operationalise contrastive
  explanations through the use of counterfactuals, by
  identifying minimal perturbations to the input that result in a change in
  the model's prediction. These minimal changes serve as the explanatory
  evidence, isolating the features most responsible for the observed
  decision. Contrastive explanations may thus also be regarded as a special
  case of feature attribution.
  
A possible instantiation of this idea within a ranking
  scenario would lead to pairwise document explanations (see
  Section~\ref{sec:ppl}). For example, if document $D_i$ is ranked higher
  than document $D_j$, a contrastive explanation might identify a set of
  ``good'' terms that contribute positively to the rank of $D_i$ because of
  their significant presence in $D_i$; these terms would typically be
  absent (or only weakly present) in $D_j$, leading to its inferior rank.

\citet{counterfactual-editing} present a concrete
  application of counterfactual reasoning in a retrieval setup. Their
  method aims to minimally perturb the \emph{query}, instead of the
  document. Given a query $Q$ and documents $D_i, D_j$ (where $D_i$ is
  ranked higher than $D_j$), their objective is to construct a $Q'$ by
  counterfactually editing $Q$, for which $D_j$ would be ranked higher than
  $D_i$. Specifically, their method first identifies important query terms
  using ColBERT term-weighting (Equation~\ref{colbert-eq}). The selected
  token is then masked and appended to the beginning of the counterfactual
  document ($D_j$). Next, DistilBERT is employed to predict replacement
  tokens, maintaining a beam of possible candidates. For each candidate, if
  the resulting retrieval score of $D_j$ surpasses that of $D_i$, the
  modified token is identified as the effective counterfactual edit;
  otherwise, additional tokens are iteratively modified until the ranking
  change is achieved. The proposed framework shows success on QA
  datasets, such as, MSMARCO, NaturalQA, FEVER, etc.
While this study represents a first attempt, it has some drawbacks.
  There is no constraint on the maximum number of query terms that can be
  modified. Thus, the edited query may actually be entirely different from
  the original. Although such extensive edits might successfully promote
  the counterfactual document in the ranking, they risk altering or
  diminishing the original information need of the user.

\citet{attacks-sigir-23} and related studies
(\cite{prada-tois-23,nrm-robust-tois-2023}) also use similar ideas
(involving document edits instead of query edits), but their focus is on
studying the robustness of NRMs under such `attacks', rather than
explanations, per se. These articles are therefore not discussed in detail
in our survey.
  


\section{Basic Methods for Constructing Post-hoc Explanations}
\label{baseline-methods}
Before moving to a detailed discussion of recent explainability research in
IR / NLP, we first review some well-established methods that are widely used
as baselines against which more recent approaches are compared. These
methods will be referred to repeatedly in the following sections and form a
good starting point for exploring this sub-field.

\subsection{Using probing classifiers to construct global explanations}
\label{sec:probing}
Probing classifiers are being widely adopted for analyzing deep neural
networks. They are based on the following idea. Let $\mathcal{M}_C$ be a neural
model, trained for a specific target task, e.g., machine translation. Given
a (preferably large, representative) set of inputs $\mathcal{I} =
\{\myellipsis{I}{1}{p}\}$, we collect the final (or an intermediate)
representation of each $I_j$ --- denoted by $I'_j$ say, --- generated
within $\mathcal{M}_C$. A simple `probing' classifier, $\mathcal{M}_S$, is trained to predict
some specific linguistic property, e.g., part of speech (POS) of
$\myellipsis{I}{1}{p}$, using $\myellipsis{I'}{1}{p}$ as input. If $\mathcal{M}_S$
(the `probe') performs well, we hypothesise that $\mathcal{M}_C$ has `learned' the
linguistic property in question by or before the layer from which the
$I'_j$s are extracted. If a probe uses only the representation
generated by the penultimate layer, it may broadly be regarded as
model-agnostic; any probe that provides insights into intermediate
representations is, naturally, model-aware.

A typical probing framework would involve, for example, running a set of
given sentences through a pre-trained 
version of BERT to generate contextual representations of the constituent
tokens. A single-layer feed forward network could then be trained to take
these embeddings as inputs, and predict the Part of Speech (POS) of
each token. If the prediction accuracy of this POS tagger is high, we may
claim that BERT's representation encodes POS information. It should be
clear from this discussion that probing is a model-aware method for
providing global, post-hoc explanations.

While a number of
studies~\citep{kohn-2015-whats,gupta-etal-2015-distributional,hupkes-rnn}
have used probing methods to evaluate hypotheses like ``neural models learn
syntactic structures'', 
\citet{10.1162/coli_a_00422} highlights the drawbacks of probing methods,
some of which are described below.
\begin{itemize}[topsep=0pt,itemsep=0pt,leftmargin=*,labelindent=4pt]
\item There are no general guidelines for selecting appropriate baselines
  for probing experiments, and by implication, for determining whether the
  new classifier performs ``well enough'' to support a particular
  hypothesis.
\item Similarly, there is no well-defined process for determining whether
  the probing classifier is simple or complex, though some studies do
  suggest reporting the accuracy complexity trade-off, or deal with
  parameter free probing.
\item The effects of the size, composition, and relationship between the
  datasets used for the original and probing tasks have not been studied
  rigorously, and need to be carefully investigated.
\item Probing frameworks may be able to associate some linguistic
  property with an intermediate representation, but it does not clearly
  establish what role, if any, that particular property played in the
  prediction of the original classifier.
\end{itemize}

\subsection{Model-agnostic, local explanations}
\subsubsection{LIME}
\label{sec:lime}
LIME~\cite{lime} is one of the earliest methods to be proposed in the
current wave of research on explainability. 
Given an input $I$, LIME generates a set $\pi_S$ of
slightly perturbed instances of $I_S$ (recall that $I_S$ denotes some
interpretable representation of  $I$: for a document, $I_S$
may be a binary term-vector), by randomly setting a small fraction
of the nonzero elements of $I_S$ to zero. For each perturbed instance, the
corresponding complex feature vector $I_C$ is also constructed; together,
these comprise $\pi_C$. A {simple} classifier, e.g., a linear model,
is trained on $\pi_S$ to mimic (and thus ``explain'') $\mathcal{M}_C$ on $\pi_C$.
No assumptions are made about $\mathcal{M}_C$. LIME is thus a model-agnostic
(Section~\ref{sec:n-or-a}) method that provides simple surrogates
(Section~\ref{sec:surrogates}) as local (Section~\ref{sec:g-or-l}) explanations.

\subsubsection{Feature attribution}
\label{sec:feature-attrib}
As discussed in Section~\ref{sec:feat-attr}, the objective of feature attribution
is to quantify the contribution of a particular feature towards the output
of $\mathcal{M}_C$ on $I$. Possibly the simplest feature attribution technique is
\emph{occlusion}~\cite{10.1007/978-3-319-10590-1_53,DBLP:journals/corr/LiMJ16a}
(sometimes referred to as the leave one out method). This estimates the
importance of a group of features by setting all these features to zero,
and measuring the resulting decrease in prediction accuracy.\footnote{This
  general idea has been used in a wide variety of contexts where effects
  have to be measured.} Since this approach only examines how the output of
the model is affected by changes in the input, it is model-agnostic. Two
other commonly-used families of feature attribution methods are discussed
below. A library implementing many of these methods is available from
\url{https://captum.ai/}, and may be used to interpret PyTorch models.

\todo{Possible to fit deeplift~\cite{deeplift} into one of the two
  following subsections?}

\paragraph{Gradient based techniques}
\label{sec:integrated-gradients}
\citet{10.5555/3305890.3306024} proposed \emph{Integrated Gradients}, a
well-known method in this class of approaches.
To quantify the importance of each feature, $I_C$ (the feature vector of an
input instance $I$) is compared to a ``baseline'' ($I^{(0)}$, say) that
corresponds to a very simple input. For example, if the input is an image,
the baseline could be a black image; when studying word
embeddings, the zero vector would be a possible baseline. We consider $m$
additional instances by linearly interpolating between $I_C$ and
$I^{(0)}$.
If $\mathbf{i}_j$ denotes the $j$-th feature or dimension of $I_C$, then its
importance (or effect on the prediction) is measured in terms of its
\emph{integrated gradient} $IG_{j} (I_C, I^{(0)})$, which
depends on two factors:
\begin{enumerate*}[~(i)]
\item 
the difference between $\mathbf{i}_j$ and its corresponding value $\mathbf{i}^{(0)}_j$ in the baseline $I^{(0)}$, ~and
\item $\displaystyle\frac{\partial f(I_C)}{\partial \mathbf{i}_j}$, the
gradient of the prediction $f$ with respect to $\mathbf{i}_j$.
\end{enumerate*}
Instead of using the value of the gradient at a single point, we use the
average of the gradients computed at the $m$ interpolation points between
$I_C$ and $I^{(0)}$. More concretely,
\[ IG_{j} (I_C, I^{(0)}) = (\mathbf{i}_j - \mathbf{i}^{(0)}_j) \times
  \frac{1}{m}\sum_{\alpha=1}^{m} \frac{\partial f(I^{(0)} + \frac{\alpha}{m}\times(I_C - I^{(0)}))}{\partial \mathbf{i}_j}.
\]


\paragraph{Shapley}
\label{sec:shapley}
The Shapley value~\cite{shapley} was originally proposed in 1952 in the
context of Cooperative Game Theory to quantify the contribution of a player
to the outcome of a game. When a team of $k$ players participates in a game
and receives a payout, 
the objective of Shapley is to distribute the payout among team members in
a \emph{fair} way, i.e., in accordance with their respective contribution.
\citet{deep-shap} first proposed the use of Shapley values to explain ML
models: each feature may be regarded as a player, the model itself is the
game, the prediction of the model is the payout, and the Shapley value for
each feature measures its contribution towards the overall prediction. The
time complexity of computing Shapley values according to the original
formulation is exponential, since it involves examining all possible
subsets of features. \citet{deep-shap} formulate approximate versions,
viz., Kernel SHAP, Tree SHAP (tree based approximation of SHAP), and Deep
SHAP, that may be computed in reasonable time.

\section{Evaluating explanations}
\label{sec:eval-expl}
The following sections describe diverse approaches that have been
proposed to explain neural models for text representation / indexing
and ranking. In order to measure advances in this area, an
  evaluation framework is crucial.
Humans constitute
the primary target of explanations; the diversity of this target group
makes it challenging to formulate an evaluation framework. The
consumers of explanations range
from lay users, through persons responsible for checking
compliance with ethical / legal requirements, and domain experts who may not be
familiar with the principles underlying IR / NLP models, to IR / NLP
researchers and practitioners, whose primary objective may be failure
analysis, or to gain insights that guide further development. 
Of course, this diversity is also true of the users of IR / NLP systems, and
the original Cranfield paradigm for evaluating IR had several
drawbacks. Some of these have been addressed, while some remain, but over
the years, a generally accepted protocol for evaluating IR / NLP systems has
emerged. Importantly, a number of large test collections have been created
for many different tasks, and their role in advancing the field cannot be
overemphasised.

Even though similar, widely accepted protocols and test
collections for ExIR are yet to emerge, two broad evaluation paradigms may
be identified. The first involves anecdotal evaluation by humans.
This seems unavoidable, since most research groups do not have the resources
to conduct carefully designed, large-scale user studies. While this sort of
evaluation may be enough to establish the plausibility of
explanation methods, it is not rigorous. 
The second approach to evaluation utilises different quantitative notions
of similarity. When explanations are in the form of simple surrogates that
mimic complex models (see Section~\ref{sec:nature-or-form}), evaluation is
typically based on measuring the similarity between the behaviours of the
explanation and the complex model, using standard metrics like rank
correlation. In the ExIR literature, these metrics are
  generally termed \emph{fidelity measures}. An explanation is considered
  \emph{faithful} if it precisely mimics the underlying model, or
  accurately captures the reasoning underlying the model's decision.

Likewise, feature attribution is generally evaluated by measuring the
difference in the performance of a model when it is given the original
inputs, and when it is given a simplified representation consisting of just
the putatively important features. In this context,
  \emph{sufficiency} and
  \emph{comprehensibility}~\cite{deyoung-etal-2020-eraser} are two metrics
  that are widely used in the explainability literature. The sufficiency
  metric quantifies the extent to which the explanation component alone results
  in an accurate prediction. In contrast, comprehensibility measures the
  degree to which removing the explanation component causes a significant
  change in the model’s output. Intuitively, a high score on either metric
  indicates that the explanation component plays an important role in the
  model’s prediction process.
Rationale-based approaches may be evaluated similarly. In cases where human
annotated rationales are available, explanations may also be evaluated by
measuring the overlap between the generated rationales and the ground
truth. These metrics are often referred to as
  \emph{correctness} measures.

  Despite the existence of these metrics, evaluation was the
  focus of a long and animated discussion at the SIGIR 2025 workshop on
  Explainability in Information Retrieval~\cite{xir-workshop-sigir-25}, and
  the lack of a definitive evaluation framework for ExIR was generally
  acknowledged. Thus, this is arguably the biggest issue facing the
  community, and needs to be resolved urgently.

\todo{complete this section}


\section{Document Ranking}
\label{sec:document-ranking}
Pre LLM era Neural models for document ranking are generally categorised as
\emph{representation-based} or \emph{interaction-based}.
Like traditional retrieval models, representation-based models also
encode queries and documents as vectors; however, in contrast to the sparse
vectors used by traditional approaches, these vectors are dense. On the
other hand, interaction-based models capture complex interactions across
query-term--document-term pairs. 
This section provides an overview of efforts to explain both categories of
neural ranking methods, but first, we provide a brief review of some
foundational attempts to understand traditional IR models
(Section~\ref{sec:traditional-ir}). These early studies were all targeted
towards \emph{global} explanations. Next, in
Section~\ref{sec:axiomatic-framework}, we discuss the Axiomatic
Framework~\citep{fang-sigir-04} --- a cornerstone of a substantial body of
work on explaining IR --- and its recent applications. In
Sections~\ref{sec:additive-models}--\ref{bert-ranking}, we review other
approaches to providing global explanations of Neural Ranking Models
(NRMs). In Sections~\ref{lime-in-ranking}--\ref{shap-listwise}, we turn to
studies devoted to local explanations. 
Tables~\ref{tab:summary-table-dataset-ranking-global} and
\ref{tab:summary-table-dataset-ranking-local} summarise the articles
covered in these sections, according to whether they address
global or local approaches. We have also prepared a Python package,
\texttt{ir\_explain}~\cite{ir-explain-resource}\footnote{Available to use from \url{https://github.com/souravsaha/ir\_explain}}, that provides
implementations of well-known post-hoc explanation strategies (including
pointwise, pairwise, and listwise methods) within a single framework.

%
%

\subsection{Understanding traditional models: the vector space model, and
  language modeling}
\label{sec:traditional-ir}
Traditional IR systems typically represent documents using \emph{terms},
i.e., single words (or word stems), and sometimes, additionally, groups of
words that have special significance (these are often loosely referred to
as \emph{phrases}), along with their associated weights. The term weights
(or feature weights) are calculated using either intuitive heuristics, or
on the basis of various mathematical models. Traditional IR models may
thus be regarded as explainable \emph{by design}.
%
Nonetheless, some early work~\citep{amit-thesis,kamps2005importance,smucker2005investigation} focused on developing a better understanding
of these models. Since these studies try to provide general
  characterisations of specific models, in terms of the criteria introduced in
  Section~\ref{sec:basic-terminology}, their explanations may be termed
model-aware and global. We summarise these early efforts below.

\citet{amit-thesis} carefully examined traditional
term-weighting formulae used in the Vector Space Model, explained why they
do not work as well as more effective statistical models like BM25,
proposed ways to modify the formulae in order to address these flaws, and
demonstrated that the modified formulae do indeed yield significant
improvements in retrieval effectiveness. Besides conventional
text collections, \citeauthor{amit-thesis} also considered noisy text
collections, such as those generated by running Optical Character
Recognition (OCR) on scanned document images, and explained why
it is inadvisable to use cosine normalisation for such collections.
\citet{kamps2005importance} studied similar issues within the
context of retrieving XML elements from a collection of XML documents,
to explain the importance of length normalisation in XML retrieval.
On a related note, \citet{smucker2005investigation} compared two well-known
smoothing techniques used in Language Modeling, and presented a different
explanation for why Dirichlet smoothing outperforms Jelinek-Mercer
smoothing, compared to the perspective provided by \citet{zhai2004study}.
%
We conclude this section by mentioning
I-REX~\citep{i-rex}, an interactive tool that is intended to facilitate
further studies along these lines by providing a white-box view
into traditional IR models. For example, given a retrieval model, and a
query $Q$, the \textbf{\texttt{compare}} command in I-REX provides a
pairwise explanation (Section~\ref{sec:ppl}) of the ranks of documents $D_i,D_j$,
by displaying the query terms present in $D_i,D_j$ along with their term
frequencies, collection frequencies, and overall contribution to the
query-document scores.

\subsection{The axiomatic framework: improving and explaining IR models}
\label{sec:axiomatic-framework}
There are certain accepted principles (or heuristics)
underlying the different term-weighting formulae that are commonly used,
even though they are derived using different mathematical models.
  For example, it has been generally observed that documents
  having multiple query term matches are more likely to be relevant, and
  should therefore be ranked above documents  
  that have a very strong match on only one term. In a series of
studies~\cite{fang-sigir-04,semantic-term-matching-fang,fang-tao-zhai-tois-11},
\citeauthor{fang-sigir-04} consolidated these principles, and formalised
them within an axiomatic framework, as a set of constraints, or axioms,
that term-weighting schemes must satisfy, in order to be effective. This
framework provides a natural basis for pairwise explanations (Section~\ref{sec:ppl}), i.e., for
reasoning about why a particular document $D_i$ is ranked above $D_j$ by a ranking
model for a given query $Q$. It has been used to explain why certain
retrieval models work better than other models, as well as to formulate
improved versions of existing retrieval
models by better aligning them with IR axioms.~\cite{log-logistic-model,clinchant2009bridging}. Later efforts have
focused on refining the framework by incorporating additional
axioms~\cite{proximity-measure-ir,semantic-term-matching-fang}, as well as
expanding it to cover pseudo-relevance feedback
methods~\cite{clinchant2013theoretical}. Table~\ref{tab:axiom-table} lists
some of the most popular axioms that have been mentioned in the literature.
The page at \url{https://www.eecis.udel.edu/~hfang/AX.html} is an excellent
resource on this topic.

\begin{table}[h]
  \centering




  \begin{tabular}{c l L{0.65\textwidth}} \toprule
    Axiom Group & Axiom Name & Axiom Statement \\\otoprule

    \multirow{6}{2cm}{\textbf{Term\lb{}Frequency}}
    & TFC1 & favour a document with more occurrences of a query term \\
    & TFC2 & ensure that amount of increase in retrieval score of a document 
    must decrease as more query terms are added \\
    & TFC3 & favour a document matching more distinct query terms \\
    & TDC  & favour a document matching more discriminating query terms \\
    & TF-LNC & regulates interaction between term frequency and document length \\\midrule

    \multirow{2}{2cm}{\textbf{Document\lb{}Length}}
    & LNC1 & penalize a long document (assuming equal term frequency)\\
    & LNC2 & over-penalize a long document \\\midrule

    \multirow{7}{2cm}{\textbf{Semantic\lb{}Matches}}
    & STMC1~\cite{semantic-term-matching-fang} & 
    favour a document containing terms that are more semantically related to query terms \\ 
    & STMC2~\cite{semantic-term-matching-fang} & avoid over-favouring semantically similar terms\\
    & STMC3~\cite{semantic-term-matching-fang} & 
    favour a document containing more terms that are semantically related to \emph{different} query terms \\
    & TSFC~\cite{Pang2017ADI} & a document having higher exact match should get higher score if their semantic similarity score matches\\\midrule

    \textbf{Term Proximity}
    & TP~\cite{proximity-measure-ir} & favour a document in which query terms appear in closer proximity \\\bottomrule 
  \end{tabular}
  \caption{Popular axioms proposed in the past literature on axiomatic IR research (Section~\ref{sec:axiomatic-framework}), adapted from~\cite{fang-tao-zhai-tois-11,camara-hauff-ecir}.}
  \label{tab:axiom-table}
\end{table}

The axiomatic framework has been applied in recent times to analyse and
understand neural IR models in terms of the common principles underlying
traditional IR models (which are, in a sense, explainable by design). Some
of these efforts (e.g., \cite{rennings-diagnostic-dataset}) investigate the extent to which various IR models (neural,
as well as traditional) satisfy axiomatic constraints. Others (e.g.,
\cite{volske2021towards}) try to
approximate the ranking generated by a
retrieval model via an explainable surrogate, a la
LIME~(Section~\ref{sec:lime}). In the rest of this section, we discuss
(in chronological order) these attempts to study NRMs using the axiomatic
framework.

\citet{Pang2017ADI} investigated the difference between the
manually engineered features used in traditional models (BM25 and LMJM) and automatically learned features derived from two pre-BERT
NRMs: ARC-I~\cite{arc-i}, a representation based model, and
MatchPyramid~\cite{matchpyramid}, an interaction based model. Their
experiments on the Robust04 and LETOR 4.0~\cite{letor}
datasets suggest that features automatically learned by NRMs are
more robust compared to the word-based features used by BM25 and LMJM, in the
sense that the performance of NRMs is less affected than that of
traditional models when some features (according to the order of feature importance score) are removed. They also confirmed that
interaction-based models appear to focus on query related words,
while representation-based models seem more concerned with general
topic related words, such as those generated by a state-of-the-art LDA
based model.
Interestingly, \citeauthor{Pang2017ADI} noted that ARC-I and MatchPyramid
performed poorly on LETOR 4.0 and even worse on Robust04, compared to
BM25 and LMJM. Their failure analysis suggests that these NRMs do not
satisfy the TDC axiom 
(Table~\ref{tab:axiom-table}). In addition, due to limitations on memory and time,
documents were sometimes truncated to some fixed length, e.g., 500 words.
This truncation may reduce the number of query terms present in a document,
resulting in a violation of the TFC1 axiom. It was also suggested that
semantic matching may be introducing too much noise into the scoring function. A
new axiomatic constraint TSFC (Table~\ref{tab:axiom-table}) was proposed, which states that if two documents are of equal length, and have
equivalent semantic term matching signals, the document having a larger
exact match should be ranked higher. Furthermore, the authors argued
that the size of Robust04 is not suitable for training NRMs.
\todo{how the words are generated in arc-i pooling e? second version e.}

Following the trend of creating diagnostic datasets for NLP and Computer
Vision applications, \citet{rennings-diagnostic-dataset} used
selected axioms to construct a similar dataset for IR.
Query-document-pair triplets that satisfy TFC1, TFC2, TDC (cf.\
Table~\ref{tab:axiom-table}) were sampled from the WikiPassageQA~\cite{wiki-passage-qa}
corpus. Artificial documents were constructed to create triplets satisfying
LNC2. This diagnostic dataset was used to evaluate the \emph{axiomatic
  efficacy} (the fraction of axiomatic constraints that is satisfied by a retrieval
model) of both traditional models (BM25 and LMDIR), and neural
models like DRMM~\cite{drmm}, aNMM~\cite{anmm}, Duet~\cite{duet} and
MatchPyramid~\cite{matchpyramid}.
The authors observed that statistical models outperform neural models
in terms of axiomatic efficacy, but were significantly
outperformed by DRMM and aNMM in terms of retrieval effectiveness. 
In later work, \citet{camara-hauff-ecir} extended this idea by
considering five additional axioms (LNC1, STMC1, STMC2, STMC3, TP). Once
again, empirical results showed that DistilBERT~\cite{Sanh2019DistilBERTAD}
significantly outperforms traditional query likelihood methods, but it does
not perform as well in terms of axiomatic efficacy. It is particularly
counter-intuitive that this observation applies to semantic matching
axioms, e.g., STMC1, as well. These results suggest that existing axioms,
including the semantic matching axioms, are not adequate for analysing NRMs
based on BERT and its variants, and need to be formulated more carefully.

\citet{volske2021towards} provided post-hoc explanations of IR systems by
constructing an \emph{explanation model} to approximate the ranking
generated by a retrieval model. Recall that each axiom induces relative
orderings for certain document pairs; e.g., axiom $A_i$ may suggest that
$D_j$ should be ranked above $D_k$. The explanation model (a random forest)
may be regarded as a weighted aggregation of these orderings.
In the words of the authors, ``(a) the explanation model's parameters reveal
the degree to which different axiomatic constraints are important to the
retrieval model under consideration, and (b) the fidelity with which the
initial ranking can be reconstructed can point to blind spots in the axiom
set, which can help to uncover new ranking properties yet to be
formalized.''
%
Twenty axioms were used in this study. 
The behaviour of three statistical models (BM25, TF-IDF, PL2) and five
neural models (MP-COS~\cite{matchpyramid}, DRMM~\cite{drmm},
PACRR-DRMM~\cite{pacrr-drmm}, BERT-3S~\cite{bert-3s},
DAI-MAXP~\cite{dai-maxp}) were studied using the MSMARCO~\cite{msmarco} and
Robust04 collections (please see Table~\ref{tab:dataset-ranking} for a
summary of information about some datasets commonly used in IR explainability studies). The authors observed that it is difficult to explain
the relative ordering of two documents if they have similar retrieval
scores, but a big difference in scores can be reliably explained in terms
of the axioms. They also noted that explanation effectiveness is comparable
for traditional and neural models (in this study, explanation efficacy is
measured in terms of how accurately the explanation model can replicate a
retrieval model's ranking), but it was higher for the Robust04 collection,
as compared to MSMARCO.

The use of the axiomatic framework in these
studies~\cite{rennings-diagnostic-dataset,camara-hauff-ecir,volske2021towards}
draws on earlier work by \citea{hagen-cikm-16} that is not directly related
to explainability. \citeauthor{hagen-cikm-16} were likely the first to
formulate \emph{relaxed} versions of axioms that permitted them to be used
for practical analysis,. Axiom TFC1 (cf.\ Table~\ref{tab:axiom-table}), for
example, induces an ordering on a document pair based on the
{term-frequencies} of a query term in the two documents, provided the two
documents are of \emph{equal} length. To be applicable in practice, this
equality condition was replaced by an \emph{approximate equality}
constraint: axiom TFC1 was applied to a document pair if the relative
difference in the lengths of the two documents was at most 10\%. The work
discussed
above~\cite{rennings-diagnostic-dataset,camara-hauff-ecir,volske2021towards}
uses such relaxed formulations of various axioms. 
%
  It is worth noting that, even after the relaxations
  mentioned above, a number of the axioms (especially the later additions)
  involve pre-conditions that do not hold for most documents (or document
  pairs) retrieved by real models from benchmark collections. Thus, these
  axioms can rarely be applied in practice. Future research should focus on
  developing more representative axioms and rigorous test suites to better
  ground axiomatic principles in practical retrieval scenarios.


\subsection{Global explanations: additive models as explainable surrogates
  or as inherently explainable models}
\label{sec:additive-models}
\citet{10.1145/3397271.3401286} studied simple, additive models as
surrogates of both statistical and neural retrieval
models, by projecting these models onto a 3-dimensional space, with
well-understood dimensions, corresponding to term frequency (\textit{tf}),
document frequency (\textit{df}) and document length (\textit{len}). For
each of the ``complex'' models, a simple, linear regression model was
trained to predict the score of a term in a document using \textit{tf},
\textit{idf} and \textit{len}. For neural models, semantic overlap was
introduced as a fourth dimension to capture the contribution of soft
(non-exact) term matches to the overall similarity between the
query and a document. The regression models are intended to provide linear
approximations (as shown in the example below) of the non-linear
expressions used by ranking models:
\begin{equation*}
  \mathit{wt}_{\mathcal{M}} (d_i, D) \approx \alpha  \times \mathit{tf}(d_i, D) ~+~ \beta \times \mathit{idf}(d_i) ~+~ \gamma \times \mathit{len}(D) ~+~ \delta \times \mathit{semantic\_overlap}(d_i, D). 
\end{equation*}
Here, $wt_{\mathcal{M}} (d_i, D)$ denotes the weight of term $d_i$ in a document
$D$ as computed by model $\mathcal{M}$, and $\delta = 0$ for traditional models. The parameters learned from the
regression model were used to ``explain'' how different ranks were produced
by four retrieval models: BM25, LMJM, LMDIR and DRMM.

Generalised Additive Models (GAMs~\cite{gam}) are, as the name
suggests, a generalisation of linear regression models. 
Instead of learning just coefficients, a GAM learns a
smooth function for each input feature.
The outputs of these functions
are simply added together to obtain the prediction $y$ for an input
instance. GAMs are regarded as
interpretable~\cite{10.1145/2339530.2339556}. 
%
\citet{neural-rank-gam} applied GAMs to the Learning to Ranking (LETOR)
problem. In their algorithm, NeuralRankGAM, each function constituting
the GAM was represented by a simple feed forward neural network,
leading to the claim that the resultant architecture is inherently
interpretable. The authors adapted piece-wise
regression~\cite{est-regression-models-unk-break-points} to construct
simpler, \emph{distilled} versions of these networks in order to make
inferencing faster, as well as to make the models potentially more
interpretable. Of course, it is debatable whether a ``simple'' model
involving ``only'' a few thousand parameters is actually interpretable.
On three LETOR datasets (Yahoo~\cite{yahoo-ltr},
Web30K~\cite{Qin2013IntroducingL4}, and a private dataset sub-sampled from
the Chrome Web Store logs), NeuralRankGAM outperformed, in terms of NDCG, other GAMs (TreeGAM~\cite{10.1145/2339530.2339556} and Tree
RankGAM~\cite{lightgbm}) that were
used as baselines. Interpretability was studied using visualisations of
occlusion-based feature attribution scores (Section~\ref{sec:feature-attrib}). The model appeared no more
interpretable than its competitors. 



\citet{ilmart} proposed another ``inherently interpretable'' ranking model
based on LambdaMART~\cite{pmlr-v14-burges11a}, that has a similar structure
as GAMs, but uses ensembles of trees. Their approach uses a modified
boosting LambdaMART algorithm in three stages. In the first stage, an
ensemble of trees is used, with each tree working on a single feature. The
interaction between distinct pairs of features (as identified in the first
stage) is modeled in the second stage. 
These features correspond to standard LETOR features, including BM25 score, document length, PageRank, etc.
In the last stage,
the LambdaMART algorithm is restricted to use only the top $k$ feature pairs obtained
from the previous stages.
The model was compared
against NeuralRankGAM (discussed above) and a pointwise ranking algorithm
EBM~\cite{ebm}, on three learning to rank datasets,
Istella-S~\cite{istella-s}, Web30K, and Yahoo. The proposed model resulted
in further improvements in NDCG, but its claim of intepretability is open
to debate.

\subsection{Global explanations based on probing}
\label{sec:ranking-probing}
Probing (see also Section~\ref{sec:probing})
attempts to characterise
linguistic features that are encoded within the representations generated
at intermediate layers of a pre-trained network. In an IR context, early
experiments with probing were reported by
\citet{deep-sentence-emb-for-lstm-to-ir} and
\citet{probing-bruce-croft-ictir-18}.
\citeauthor{probing-bruce-croft-ictir-18} trained a multilayered LSTM
network for an IR-oriented task, viz., answer passage retrieval for the
CQA~\cite{surdeanu-etal-2008-learning} and
WikiQA~\cite{yang-etal-2015-wikiqa} datasets. 
After training, the model weights were frozen, and the intermediate representations were subsequently used for probing—that is, to investigate what linguistic or semantic information was captured at different layers of the network, following a strategy similar to that proposed by~\citet{alain2017understanding}.
At a high level, \citet{alain2017understanding} employed linear classifiers to examine the extent to which internal representations within an image classification network encode specific forms of information. After training the network, they froze the weights and used the hidden-layer representations as features for simple linear classifiers to assess the presence of particular attributes.
Building upon this idea, \citet{probing-bruce-croft-ictir-18} extended the probing methodology to several NLP tasks, including part-of-speech (POS) tagging on the Penn Treebank~\cite{penn-treebank}, named entity recognition (NER) on CoNLL-2003~\cite{conll}, sentiment classification on the IMDB review dataset~\cite{maas-etal-2011-learning}, and textual entailment using the Stanford Natural Language Inference (SNLI) corpus~\cite{snli-corpus}.
\citeauthor{probing-bruce-croft-ictir-18} found that initial layers appear
to capture NER and POS features, while the later (second LSTM) layer is
concerned with higher-level, abstract topical representations.
Interestingly, the study by \citeauthor{deep-sentence-emb-for-lstm-to-ir}
indicates that such conclusions may be collection-dependent, i.e., the
abstract representations appear to capture different features for different
collections. For example, representations from the WikiQA model are more
effective for POS tagging and NER than those from the CQA model, but they
perform poorly on the sentiment classification and entailment tasks.

\citet{Qiao2019UnderstandingTB} highlighted the interaction-based,
sequence-to-sequence nature of BERT using a few different BERT-based
rankers. A representation-based BERT model was designed by taking the
individual [CLS] embeddings for both queries and documents, and computing
the cosine similarity between them. Experiments on the
MSMARCO passage reranking and TREC Web Track 
document ranking tasks
showed that, on both datasets, the representation-based BERT model
performed poorly. This suggests that BERT should not be regarded as a representation-based model.
Interestingly, \citeauthor{Qiao2019UnderstandingTB} observed that, while
monoBERT~\cite{DBLP:journals/corr/abs-1901-04085} outperformed the
then-state-of-the-art NRMs (K-NRM, Conv-KNRM) on MSMARCO, it
failed to perform well on document ranking.
They concluded that passage ranking for MSMARCO is more related to a
question answering (QA) task, and therefore, better suited for a seq2seq
model. Indeed, MSMARCO is basically a QA dataset, where precise matches are
important. In contrast, traditional IR datasets,
where recall is also important were used to train Conv-KNRM. It would be interesting to fine-tune BERT
on such datasets, and then compare its performance against
Conv-KNRM.
%
On the other hand, their experiments showed that removing a single
term from a document affects its score as calculated
by the Conv-KNRM model, but monoBERT is more robust to such changes.
%
\citeauthor{Qiao2019UnderstandingTB} also contributed to the ongoing debate
about attention (this is summarised in Section~\ref{debate-on-attention}). They
observed that the [CLS] and [SEP] tokens received the most attention.
Further, even though removing stopwords does not affect the effectiveness
of a ranker, stopwords and non-stop words received similar attention scores.

A more recent series of experiments by~\citet{10.1162/tacl_a_00457} uses a different form of probing to analyse the sensitivity of
NRMs to various changes in textual properties. 
In a sense, \citeauthor{10.1162/tacl_a_00457}'s approach is similar in
spirit to feature-attribution studies: it looks at
how the output changes when changes are made to input features. However,
their objective is to obtain a global characterisation of NRMs, rather than
determine which features of a particular input instance are important.
A given NRM (trained on the
MSMARCO passage ranking task) was first probed using a query-document pair
$\left\langle Q,D \right\rangle$, and then using $\left\langle Q,D' \right\rangle$, with $D'$ being
a perturbation of $D$. Four broad classes of perturbations were studied:
\begin{enumerate*}[~(a)]
\item ``Measure and Match'', in which $D'$ was generated by varying a
  specific attribute of $D$ (e.g., term frequency), while keeping other
  attributes (e.g., document length) fixed; 
\item simple textual manipulation (changing the order of
  words in a passage, introducing typographic errors, etc.);
\item ``Dataset Transfer Probes'' (DTPs), which involve more complex,
  stylistic changes, e.g., changes in the level of fluency, formality, or
  the succinctness of a passage; and
\item changes in factual content, which typically involve replacing one
  entity (e.g., New York) by another entity of the same type (e.g., Moscow).
\end{enumerate*}
For the first two types of perturbations, synthetic documents were created,
while DTPs were created from datasets specifically designed for the purpose
(e.g., summarisation datasets). 
If a perturbation resulted in significantly different scores for $D$ and
$D'$, the corresponding NRM was characterised as being
sensitive to that perturbation.
%
Experiments were conducted on the TREC-DL 19 passage ranking
data and ANTIQUE~\cite{antique}. A wide range of
models were studied: BM25, WMD~\cite{word-mover-dist},
SBERT~\cite{reimers-gurevych-2019-sentence}, LGBM~\cite{lightgbm},
VBERT~\cite{DBLP:journals/corr/abs-1901-04085}, T5~\cite{JMLR:v21:20-074},
EPIC~\cite{epic}, DT5Q~\cite{Cheriton2019FromDT},
ANCE~\cite{ance}, and ColBERT~\cite{colbert}. Empirical
observations include the following.
\begin{itemize}[topsep=0pt,itemsep=0pt,leftmargin=*,labelindent=4pt] 
\item Traditional models are better than NRMs at handling
  typographical errors artificially introduced into the text.
\item When non-relevant content is added to a
  document, its BM25 score generally decreases, but the score computed by
  contextualized NRMs often \emph{increases}.
\item Most contextualized models work well on full texts, while BM25 has a
  strong affinity towards summaries.
\item Some neural models, viz., ColBERT, T5 and WMD, appear to ``memorise''
  facts from the corpora they are trained on, and generally assign lower scores
  to factually incorrect passages (e.g., ``New York is the capital of
  Russia'').
\end{itemize}

We now discuss the probing framework used to analyse complex neural models, particularly large language models (LLMs). Low-Rank Adaptation (LoRA)~\cite{hu2022lora} has emerged as an efficient alternative to full fine-tuning of LLMs, offering comparable performance at a fraction of the computational cost. LoRA achieves this by freezing the pre-trained weights of an LLM and introducing trainable low-rank matrices into the Transformer architecture.

LoRA-based adaptation of LLMs for passage ranking on MS MARCO demonstrates promising retrieval performance. \citet{james-allan-short-probing} examined how relevance (i.e., performance on retrieval tasks) is encoded across LLM layers, showing that LoRA updates to multi-head attention outperform those to multi-layer perceptron (MLP) layers, while jointly updating both further enhances ranking effectiveness across three LLMs: Llama 3.1–8B, Mistral–7B, and Pythia–6.9B. Notably, the study also found that mid-level MLP layers (layers 5–15 out of 32) contribute most to relevance prediction, emphasising the layer-specific nature of retrieval signals in LLMs. However, the reported ranking performance was evaluated only on the top-$10$ retrieved documents, whereas the average number of relevant documents per query in the TREC DL 2019 dataset is approximately $95$, suggesting a limited evaluation depth.

In a subsequent study, \citet{probing-llm-ictir-25} extended this line of work to analyse what types of features are encoded across different MLP layers. Specifically, they used four LLMs—Llama-2-7B, Llama-2-13B, Llama-3.1-8B, and Pythia-6.9B—each LoRA fine-tuned on the MS MARCO passage ranking dataset. Experimental results were reported on TREC DL 2019, TREC DL 2020, and the MS MARCO development query set. They examined a range of statistical features (e.g., average term frequency, maximum term frequency, BM25 score, and document length) alongside semantic features such as various distance-based query–document interaction measures. This analysis revealed that ranking-oriented LLMs also rely on term importance and exact matching signals, showing similar feature patterns across all four models. As in previous studies, while complex LLMs capture intricate relevance relationships, BERT and T5-based semantic features appear to be less prominently encoded. Interestingly, the analysis further showed that the final layer of the LLM encodes a linear combination of three simple features—query term frequency, variance of TF-IDF, and document length-normalised term frequency (\emph{tf})—indicating that a composite of these basic features plays a crucial role in modelling relevance.

\subsection{Global explanations for BERT-based NRMs: what knowledge do
  they possess?}
\label{bert-ranking}
Instead of probing, some studies have examined attention maps and
query-document interactions, across different layers and heads, to
understand what sort of knowledge BERT-based models encode internally that
other statistical and older NRMs do not.

\citet{10.1145/3397271.3401325} studied the monoBERT model focusing on both
its attention mechanisms and query-document interactions. Once again, the
MSMARCO passage ranking task was used for experiments. As in
\cite{Qiao2019UnderstandingTB} (discussed above) and similar studies
conducted within the NLP
community~\cite{clark2019what,rogers-etal-2020-primer,kovaleva-etal-2019-revealing},
empirical results show that BERT puts redundant attention weights on [CLS]
and [SEP] tokens. 
Query-document interactions were found to play a major role in document
ranking. Their observations regarding the behaviour of different layers of
BERT echo earlier studies: interactions are not important in the initial
layers of BERT, where it appears to learns more context specific
information; in the later layers, it heavily uses interactions between a
query and document to predict relevance. These findings were confirmed by
experiments with probing
(Section~\ref{sec:probing}) and attribution (Section~\ref{sec:feature-attrib}) techniques.

Recently, ColBERT has gained popularity in the IR community. Its scoring
function is as follows:
\begin{equation}
  \mathit{Score}_{\mathrm{ColBERT}} (Q, D) = \sum_{q_i \in Q} \max_{d_j \in D} E_{q_i} E_{d_j}^{T}
  \label{colbert-eq}
\end{equation} 
where $E_{q_{i}}$ and $E_{d_{j}}$ denote the embeddings of the $i$-th query
term and $j$-th document token, respectively. This scoring
function is superficially similar to early embedding-based similarity
functions, except that $E_{q_{i}}$ and $E_{d_{j}}$ are heavily
contextualised representations of query / document tokens.
In earlier neural IR models, like DRMM~\cite{drmm}, term heuristics (e.g.,
inverse document frequency) were explicitly involved in the architecture;
ColBERT eliminates the need for such measures.
\citet{10.1007/978-3-030-72240-1_23} present a ``white-box analysis'' of
ColBERT on the TREC-DL
2019 and 2020 passage retrieval tasks. For each query term $q_i$, the authors analysed the
distributions of exact matches and soft (non-exact) matches between $q_i$
and document tokens. They observed that exact matches are more important for
high-IDF query terms. This is not unexpected, since such terms are likely to be specific, and therefore
unlikely to have high semantic overlap with other terms in the document.
The authors also estimated the impact of each $q_i$ on ColBERT's
ranking using a leave-one-out study: if the correlation
($\tau_{\mathit{AP}}$~\cite{rank-correlation}) between the ranked lists for
$Q$ and $Q \setminus \{q_i\}$ is high (close to 1, say), then $q_i$ is not important.
Experiments show that ColBERT gives more importance to higher IDF terms, as
expected.
\begin{table}[ht]
  \centering
\resizebox{1\textwidth}{!}{

  \begin{tabular}{l L{0.28\textwidth} L{0.25\textwidth} L{0.25\textwidth}}
    \toprule
    Approach &  Explanation type &  Dataset(s) & Codebase  \\ 
    \toprule
    \citet{fang-sigir-04} & Proposed axiomatic framework & TREC678 & NA \\
    \citet{hagen-cikm-16} & Relaxed axioms and used for re-ranking & Clueweb09b & \url{github.com/webis-de/ir_axioms} \\
    \citet{Pang2017ADI} & Finding interaction vs representation focused NRMs & Robust, LETOR 4.0  & NA  \\
    \citet{rennings-diagnostic-dataset}	& Diagnostic datasets with axioms & Diagnostic dataset created from WikiPassageQA & NA  \\ 	
    \citet{camara-hauff-ecir} & Diagnostic datasets with axioms & Diagnostic dataset (proposed) & \url{github.com/ArthurCamara/bert-axioms}  \\ 
    \citet{10.1145/3397271.3401286}  & Additive models & MSMARCO passage & NA \\
    \citet{neural-rank-gam} & GAMs & EB30K, YAHOO, Chrome Web Store (private data) & \url{github.com/tensorflow/ranking/releases/tag/v0.3.1} \\
    \citet{ilmart}  &  Ensemble of trees & WEB30K, YAHOO, Istella-S  &  \url{github.com/veneres/ilmart} \\
    \citet{probing-bruce-croft-ictir-18} & Probing complex rankers &  CQA, WikiQA & NA  \\
    \citet{Qiao2019UnderstandingTB} & Understanding BERT for ranking & MSMARCO passage, Clueweb09b & NA \\
    \multirow{4}{0.28\textwidth}{\citet{10.1162/tacl_a_00457}} & \multirow{4}{0.2\textwidth}{Probing NRMs} & MSMARCO passage and created MMPs, TMPs from TREC-DL 2019 and MSMARCO passage  &  \multirow{4}{0.25\textwidth}{\url{github.com/allenai/abnirml}} \\
    \citet{james-allan-short-probing} &  \multirow{2}{0.25\textwidth}{Probing LLM-ranker} & MSMARCO passage & NA \\
    \citet{probing-llm-ictir-25} &  &  MSMARCO passage, TREC-DL 2019, 2020 & \url{github.com/TaKneeAa/ProbingRankLlama}  \\
    \citet{10.1007/978-3-030-72240-1_23} & Understanding ColBERT & TREC-DL 2019, 2020 passage & NA  \\
    \bottomrule
  \end{tabular}
  }
  \caption{List of global ExIR methods covered in
    Sections~\ref{sec:axiomatic-framework}--\ref{bert-ranking},
    along with datasets used, and links to codebases, where
    available.}
  \label{tab:summary-table-dataset-ranking-global}
\end{table}

\subsection{Local explanations: LIME and SHAP inspired approaches}
\label{lime-in-ranking}
We now turn to studies that propose methods for constructing local
explanations. As discussed in Section~\ref{sec:lime}, LIME based local,
post-hoc explanations have become popular for classification tasks.
Following~\citep{lime}, \citet{10.1145/3331184.3331377} proposed LIRME, an
adaptation of LIME that provides \emph{pointwise} explanations for
document ranking. $\mathcal{M}_C(Q,D)$, the score that an IR model $\mathcal{M}_C$
computes for a query-document pair $(Q, D)$, is explained by a vector of
weights corresponding to different tokens, where the weight for a token
indicates its contribution to $\mathcal{M}_C(Q,D)$. To estimate this \emph{explanation
  vector}, perturbed versions of $D$ were generated by selecting subsets of
words from $D$, and the effects of the perturbations on the retrieval score
assigned by $\mathcal{M}_C$ were noted.
Experiments were conducted on the TREC-8 adhoc collection. The authors also
proposed two metrics to evaluate the generated explanations:
\emph{consistency} and \emph{correctness}. Consistency measures how stable
the explanation vector is across perturbations, while correctness
quantifies whether an explanation vector assigns higher weights to
``relevant'' terms. Perhaps the biggest flaw of LIRME is the overly simplistic
definition of ground truth about which terms are actually relevant: this is
determined based on the weights of terms occurring in relevant documents,
calculated using a simple statistical model (LM-JM).

\citet{10.1145/3289600.3290620} also incorporated ideas from LIME to create
pointwise explanations. Their method, EXS, uses a linear SVM as the simple
model that approximates an arbitrary, black-box ranker. As usual, perturbed
versions of a document $D$ are constructed by deleting random words from
random positions in $D$. The authors mainly focus on three different ways
to estimate the relevance of a perturbed instance of $D$:
\begin{enumerate*}[~(a)]
\item \emph{top-k binary}: the perturbed version is regarded as relevant if its
  score, as computed by the simple model (linear SVM), is more than that of
  the $k^{th}$ document in the original ranked list, and is regarded as
  non-relevant otherwise;
\item \emph{rank based}: this is similar to the above strategy, except that the
  relevance score for a relevant document decays with its rank, instead of
  being 1; and
\item \emph{score based}: In this method, relevance scores are real-valued, rather
  than binary and are determined by the relative difference in score
  between the perturbed document and the top-ranked document.
\end{enumerate*}
The form of an explanation is similar to that for LIRME: it consists of a
graph displaying which words are strong indicators of relevance for a
particular query. EXS was used to explain neural
retrieval models such as DRMM~\cite{drmm} and
DESM~\cite{10.1145/2872518.2889361}, on traditional news collections used
at TREC (AP, LATIMES, Robust04 and Financial Times). Our study~\cite{ir-explain-resource} involving LIRME and EXS suggests that the explanations generated by these approaches are often unstable; even minor perturbations in the document content can lead to substantial changes in the resulting explanations.

\label{deep-invest-neuir}
\citet{10.1145/3331184.3331312} explored the use of Shapley
values (Section~\ref{sec:shapley}) for model-introspective explanations for IR. They
used DeepSHAP, which internally computes the Shapley
values of terms present in a document (a high value indicates a
bigger contribution to the prediction of relevance).
The authors compared various approaches for creating a \emph{reference
document}, in order to determine which reference document provides the best
explanations for retrieval models. For example, reference documents were
constructed by averaging the embeddings of either all words in the
vocabulary or a set of words randomly sampled from the collection,
excluding the top 1000 documents retrieved by the query.
These explanation strategies were investigated on three NRMs: DRMM, MatchPyramid~\cite{matchpyramid}, and
PACRR-DRMM~\cite{pacrr-drmm}. LIME based explanations were used as a
baseline. Although these models use standard neural network architectures,
empirical results showed that no single reference document yields the best
explanations across different neural ranking models. Further, the
performance of these retrieval models is also sensitive with respect to
different reference inputs.

\subsection{Extracting and evaluating alignment rationales as local explanations}
\label{textual-align}
As discused in Section~\ref{sec:feat-attr}, a \emph{rationale} is a set of
tokens in an input $I$ that best {explains} a model's decision for $I$.
In an IR context, \emph{alignment rationales}, which establish a
correspondence between portions (or \emph{span}s) of
a query and a document, are particularly useful. Consider the hypothetical
query \emph{\underline{Where is} SIGIR
  2026?}~\cite{textual-alignment-sigir22}, in response to which a neural
model may plausibly retrieve the span ``\emph{SIGIR 2026 will be held in
  \underline{Melbourne}}''. The underlined parts of the query and the document
constitute an alignment rationale (AR), since \emph{Melbourne}, a location, is
a response to the \emph{Where is} portion of the query.\footnote{Clearly,
  for conventional IR models, alignment rationales are limited to term
  matches between the query and the document.}
%
\citet{textual-alignment-sigir22} proposed various perturbation-based
metrics to evaluate how faithfully ARs reflect a model's behaviour.
Deletions and substitutions were considered as possible perturbations.
Empirical findings with a BERT-based ranker (DAI-MAX~\cite{dai-maxp}) that
was trained on the MSMARCO document ranking task suggest that metrics based
on substitutions are more reasonable compared to those based on deletions
for measuring faithfulness.

\citet{extractive_explanation_tois} proposed a rationale-based `interpretable' ranking algorithm. The
\emph{top} 20 sentences were extracted from a document as a rationale; a
`sparse' representation consisting only of these sentences was used to rank
each document. Two sentence-scoring approaches were explored:~
\begin{enumerate*}[(a)]
\item an end-to-end gradient-based optimisation technique that chooses
  sentences in order to maximise retrieval effectiveness in the ranking
  phase; and~
\item choosing sentences deterministically, e.g., based 
  on their \emph{tf-idf} scores with respect to the query.
\end{enumerate*}
For ranking, a monoBERT-style architecture was used with the query and the
document rationales. The system's retrieval effectiveness on the TREC-DL,
Core 17, Clueweb09, and BEIR~\cite{thakur2021beir} datasets was at par with
state-of-the-art models (BERT-3S~\cite{bert-3s},
Doc-labeled~\cite{dai-maxp} (DAI-MAXP),
BERT-CLS~\cite{DBLP:journals/corr/abs-1901-04085} (monoBERT)).
Additionally, the quality of the rationales was evaluated by humans.
\citet{jacovi-goldberg-2021-aligning} criticised this framework in the context of NLP tasks, arguing that the selected rationales may be uninformative from a user’s perspective. We observe a similar phenomenon in retrieval tasks, where the algorithm tends to highlight sentences that humans do not consider important.



\subsubsection*{Evaluating rationales}
\citet{dg-ecir-24} proposed that the quality of rationales be evaluated
using \emph{intrinsic} and \emph{extrinsic} measures. Intrinsic
evaluation quantifies the `consistency' (correlation) between the ranks of
the top-$k$ documents when scores are computed using 
the full-text vs.\ only rationales (specifically, the
concatenation of the top $m$ rationales, where rationale scores were
computed using occlusion).\footnote{In occlusion (Section~\ref{sec:feature-attrib}),
  document segments are masked in turn; the resultant changes in the
  document's score indicate the importance of the respective segment.} On
the other hand, extrinsic evaluation quantifies how well the top $m$
rationales 
overlap with the relevant passages of the documents. 
Relevant passages were identified by aligning the MSMARCO passage and
document datasets via the ``QnA'' version of MSMARCO. Experiments with these
datasets suggest that the retrieval effectiveness of a ranking model (both
BM25 and NRMs e.g., ColBERT)
does not correlate with the proposed intrinsic or extrinsic effectiveness
measures.

\subsection{Local, listwise explanations: using query expansion and query aspects}
\label{sec:qe-for-listwise}
Query expansion (QE) based approaches (see Section~ \ref{sec:surrogates})
approximate a complex retrieval algorithm with a simpler one, by adding
`explanation' terms to the original query.
IntentEXS~\cite{singh-fat-20} and Multiplex~\cite{lijun-ecir-23} are both
examples of such approaches, and broadly follow a common framework. A
ranked list is viewed as a set of preference pairs of documents. The goal
is to identify a set of explanation terms to add to the
original query, such that a simple IR model (e.g., BM25) ranks most
preference pairs in the same order as a given
NRM. 
While IntentEXS uses a greedy preference pair coverage algorithm to
maximise the above objective, Multiplex uses the modern Geno
method~\cite{geno} for optimisation.

IntentEXS used LMJM as the simple ranker to explain three pre-BERT neural
models: DRMM~\cite{drmm}, P-DRMM~\cite{pacrr-drmm},
M-Pyramid~\cite{pang2016study}. The effectiveness of the explanation terms
was evaluated using two metrics. \emph{Fidelity}, or predictive accuracy,
was measured using the correlation (Kendall's $\tau$) between the ranked lists
of documents retrieved by LMJM and the NRMs. Next, expansion terms
generated using RM3~\cite{10.1145/383952.383972},
EMB~\cite{diaz-etal-2016-query} or DESM~\cite{10.1145/2872518.2889361} were
regarded as ground truth, and \emph{descriptive} accuracy was measured
in terms of the overlap between the explanation terms and this `ground truth'.
Experiments were conducted on Clueweb09b~\cite{callan2009clueweb09}.
When the entire ranked list was considered, Kendall's $\tau$ was found to be
$< 0.5$ for the three NRMs. When only the top $10$ documents were
considered, a $\tau$ greater than $0.5$ $(0.537)$ was observed for only P-DRMM. Given
these low $\tau$ values, this method does not seem to adequately explain the
workings of NRMs. The authors showed, however, that the framework can be
used to identify model and temporal biases. For instance, in response to
the query \emph{`fidel castro and electoral college 2008 results'}, DRMM
considers documents related to events in 2004 or later. Similarly, for the
query \emph{`how to find mean'}, DRMM correctly captures the statistical
intent behind the query, whereas DESM fails to do so.
Multiplex, on the other hand, attempted to explain three NRMs (DRMM,
monoBERT and DPR~\cite{dpr}) using a combination of three simple rankers:
i) term matching (BM25), ii) position aware, and iii) semantic similarity.
A fidelity measure (the extent to which preference pairs are in
agreement with the original ranking) was the evaluation metric.
Not all pairs were considered, however; evaluation was limited to a random
sample of preference pairs for which the rank difference was large.
Within this restricted setup, Multiplex achieved a high fidelity score on
Clueweb09b and TREC-DL 2019.


\begin{table}[b]
  \centering
  \resizebox{1\textwidth}{!}{
  \begin{tabular}{l L{0.18\textwidth} L{0.25\textwidth} L{0.3\textwidth}}\toprule
    Approach &  Explanation type &  Dataset(s) & Codebase  \\\toprule 

    LIRME~\cite{10.1145/3331184.3331377} & \multirow{2}{0.22\textwidth}{Adaptations of LIME} &  TREC-8 & \url{github.com/gdebasis/irexplain} \\
    EXS~\cite{10.1145/3289600.3290620} 	& & TREC678, Robust & \url{github.com/GarfieldLyu/EXS}  \\\hline
    \citet{10.1145/3331184.3331312} & SHAP-based & Robust & No URL provided \\
    \citet{textual-alignment-sigir22}  &  Evaluating rationale faithfulness &  MSMARCO docs; TREC-DL 2019 queries & \url{github.com/youngwoo-umass/alignment_rationale} \\
    \citet{extractive_explanation_tois} & Ranking based on rationales & TREC-DL 2019, Clueweb09b, Core 17, BEIR benchmark & No URL provided \\
    \citet{dg-ecir-24} & Extracting + evaluating rationales &  MSMARCO (pass.\ \& docs); TREC-DL 2020 queries & \url{github.com/saranpandian/XAIR-evaluation-metric}  \\\hline

    \citet{singh-fat-20} & \multirow{4}{0.22\textwidth}{Explanation terms / expanded queries} & Clueweb09b  & \url{github.com/GarfieldLyu/EXS}  \\
    Multiplex~\cite{lijun-ecir-23} &  & Clueweb09b, MSMARCO passage and TREC-DL 2019 queries & \url{github.com/GarfieldLyu/RankingExplanation} \\\hline

    \citet{listwise-expl-sigir-22} & Diversity-aware listwise explanation & Wiki, MIMICS  & \url{github.com/PxYu/LiEGe-SIGIR2022}  \\ 
    
    BFS~\cite{llordes-sigir-23} & Listwise explanation & MSMARCO passage and TREC-DL 2019 queries & \url{github.com/micllordes/eliBM25} \\\hline

    RankSHAP~\cite{rankshap-DBLP:conf/iclr/ChowdhuryZA25} & \multirow{4}{0.2\textwidth}{Adaptations of SHAP for listwise ranking} &  MSMARCO (randomly selected 250 Qs), Robust04 & No URL provided  \\ 
    
    RankingSHAP~\cite{rankshap-sigir-25} &  & Synthetic, LETOR 4.0 & \url{https://github.com/MariaHeuss/RankingShap} \\
    \bottomrule
  \end{tabular}
  }
  \caption{List of local ExIR methods covered in
    Sections~\ref{lime-in-ranking}--\ref{shap-listwise}, along
    with datasets used, and links to codebases, where available.}
  \label{tab:summary-table-dataset-ranking-local}
\end{table}

\citet{listwise-expl-sigir-22}
also consider the problem of providing 
a listwise explanation of documents.
Strictly speaking, however, their
objective is not to \emph{explain} why a given list of documents was
retrieved, or ranked in the given order; it would be more appropriate to
term their ``explanations'' as \emph{annotations}. These annotations are
intended to make it easier for a user to consume the information contained
in the list of documents.
\citeauthor{listwise-expl-sigir-22} provide two types of annotations. Given
a (relatively broad) query involving multiple aspects, the first type of
annotation ``explains'' which aspects of the query are covered by each
document in the ranked list. The second type of annotation highlights the
\emph{novel} aspects covered by a particular document, i.e., the aspects
that are not already covered by documents ranked above it. The annotations
were created by modifying the encoder and decoder components of a
Transformer network. Specifically, inter-document and intra-document
attention layers were added.

Experiments were conducted on a newly constructed Wikipedia corpus, and the
MIMICS~\cite{mimics} dataset. The title of a Wikipedia article was
considered as a query. Each section of the article was treated as a unit of
retrieval (i.e., as a single document). The heading of any section was
treated as the ground-truth~/ gold-standard explanation that specifies
which aspects of the query are covered by the corresponding document. However, as
Wikipedia articles are topically organised, most sections cover
only a single aspect of a query. Therefore, documents covering multiple, overlapping
query aspects were artificially created. Given a Wikipedia
article, three sections were randomly sampled from it (without replacement)
at a time. Two documents were created from these three sections such that
the two documents had one section in common. Thus, from an article
containing $n$ sections, $\lfloor n/3 \rfloor \times 2$ documents were
created. The second dataset (MIMICS) is a search clarification dataset with
the search queries sampled from Bing query logs. For each query, the clarifying
questions are considered as query aspects.

A group of models from search result explanation tasks,
TextRank~\cite{mihalcea-tarau-2004-textrank}, a topic sensitive PageRank
variant of TextRank, LIME based explanations for monoBERT,
KeyBERT~\cite{grootendorst2020keybert}, KeyBERT with maximal marginal
relevance~\cite{mmr-sigir}, GenEx~\cite{DBLP:journals/corr/abs-2111-01314},
HistGen~\cite{outline-generation}, NMIR~\cite{nmir} and
BART~\cite{lewis-etal-2020-bart} were used as baselines. The quality of
explanations was evaluated with BLEU and ROUGE scores. Additionally,
BERTScore~\cite{Zhang*2020BERTScore:} was computed to measure the semantic
similarity between the generated explanations and ground truth
explanations. A divergence based metric, Div, was also employed.
The proposed architecture outperforms the 
baselines in terms of the above metrics.
\todo{cut down kora jay, too much lekha hoyechge}

\citet{llordes-sigir-23} proposed a more traditional, QE-based explanation
algorithm, BFS, which uses Jaccard and rank-biased overlap (RBO~\cite{rbo})
to quantify the similarity between the ranked lists produced by an NRM and
another, produced using the expanded query by
traditional ranker, such as BM25. BFS is a heuristic search algorithm that
explores the best set of terms sampled from pseudo-relevance feedback from
documents.
Intuitively, if adding a term increases the overlap between the two
ranked lists, it explores that combination. Experimental results on TREC-DL
2019 data for popular NRMs, such as ColBERT and ANCE~\cite{ance}, were
modest, with BFS achieving a maximum RBO of around 0.5 for the top 10
documents. Additionally, the running time of BFS is also high.

\subsection{SHAP inspired listwise explanations}
\label{shap-listwise}

Recently, two independent research groups have applied SHAP for listwise explanation by extending the Shapley value framework to handle feature attribution over ranked lists. Intuitively, a feature is considered important for a ranking if its removal substantially alters the order of documents. 
The general framework is based on this idea and variations found in \citet{rankshap-DBLP:conf/iclr/ChowdhuryZA25} (RankSHAP) and \citet{rankshap-sigir-25} (RankingSHAP) is summarised in the below Table~\ref{tab:shap-table}. 


\begin{table*}[htbp]
  \centering
  \begin{tabular}{l L{0.35\textwidth} L{0.35\textwidth}}\toprule
    Approach & RankSHAP  &  RankingSHAP \\\toprule 
    Ranking objective & NDCG & Kendall's $\tau$  \\
    Datasets & MSMARCO (randomly selected 250 test queries), Robust04 & Synthetic, LETOR 4.0 \\
    Models & BERT, T5, Llama2  & LGBM \\
    \bottomrule
  \end{tabular}
  \caption{RankSHAP vs RankingSHAP}
  \label{tab:shap-table}
\end{table*}

Additionally, RankSHAP formalised four axioms tailored to the ranking objective of Shapley values:
\begin{itemize}
\item Rank-efficiency — the total feature attribution for a ranking equals the gain in RankSHAP computed with all features relative to a baseline;
\item Rank-missingness — if adding a feature does not alter the ranking order, its attribution is zero;
\item Rank-symmetry — features yielding identical effectiveness scores (as measured by changes in NDCG or $\tau$) receive equal attribution; and
\item Rank-monotonicity — if adding a feature improves effectiveness in one ranking, its attribution should be higher for that ranking.
\end{itemize}

Theoretical analysis shows that RankSHAP satisfies these axioms more consistently than RankingSHAP. Empirically, RankSHAP achieves higher fidelity—that is, the extent to which top-ranked features can reconstruct the original ranking when linearly combined with their feature values—consistently outperforming RankingSHAP across datasets and models. 
Furthermore, user studies reveal stronger alignment between RankSHAP’s attributions and human judgements. However, the authors do not report statistical significance testing to support these findings, which limits the robustness and generalisability of their conclusions.



\begin{table}[h]
  \centering
  \begin{tabular}{l  c  c  c  c}
    \toprule
    \textbf{Dataset} &  \textbf{Collection Type} &  \textbf{\#Docs} & \textbf{\#Queries} & \textbf{Pool Depth}  \\ 
    \toprule
    MSMARCO passage (train) & \multirow{3}{*}{Question answering} & \multirow{3}{*}{8,841,823} & 502,939 & Shallow \\
    MSMARCO passage (dev)  & & & 6,980 & Shallow \\
    MSMARCO passage (test) 	& & & 6,837 & Shallow \\ 	
    \hline
    MSMARCO document (train) & \multirow{3}{*}{Web} & \multirow{3}{*}{3,213,835} & 367,013 & Shallow \\ 
    MSMARCO document (dev)  & & & 5,193 & Shallow \\
    MSMARCO document (test)  & & & 5,793 & Shallow \\	
    \hline
    TREC DL passage 2019~\cite{trec-2019-overview}  & Question answering & 8,841,823 & 43 & Deep \\
    TREC DL document 2019~\cite{trec-2019-overview} & Web & 3,213,835 & 43 & Deep \\
    \hline
    TREC DL passage 2020~\cite{trec-2020-overview} & Question answering & 8,841,823 & 54 & Deep \\
    TREC DL document 2020~\cite{trec-2020-overview} & Web & 3,213,835 & 45 & Deep\\
    \hline
    Robust  &  News & 528,155  &  249  & Deep\\
    \hline
    Clueweb09b  &  Web &  50,220,423 &  200 & Deep\\
    \bottomrule
  \end{tabular}
  \caption{Some statistics about datasets commonly used for
    experiments related to the interpretability of document ranking tasks.}
  \label{tab:dataset-ranking}
\end{table}

\section{Explaining Conversational and Retrieval Augmented Generation (RAG) Systems}
\label{sec:explain-rag}
As mentioned in the Introduction, conventional document
  retrieval systems are rapidly giving way to interactive, dialogue-driven
  information access. However, Conversational Information Seeking
  (CIS)~\cite{generate-clarifying-ques,lei-zamani-conversation,fntir-conv-ir}
  was an area of active research even before the recent popularity of
  interfaces driven by Large Language Models. Because they eschew a
  conventional interface involving full-text document retrieval,
  explanations are also particularly important for conversational IR
  systems. To promote transparency in such dialogue-oriented information
  access systems, \citet{explainable-conversational-ir} identified three
  key types of explanations: (i) the sources from which the system derives
  its answers, (ii) the confidence score associated with the generated
  response, and (iii) the limitations of the generated answer. They
  conducted extensive user studies using queries drawn from the TREC
  CAsT~\cite{trec-cast} collection. The results revealed that users often
  struggled to recognise misleading or noisy explanations regarding the
  limitations of generated answers. However, when a mismatch arose between
  their own confidence and the system’s stated confidence in a response,
  users were generally able to identify the inconsistency.

The emergence of LLMs has accelerated the trend towards searching via
chat-based interfaces. These LLMs are trained on massive, but static, text
corpora. As new information is created, it is important to ensure that the
responses of LLMs also remain up to date. Towards this end,
these models are generally augmented with a \emph{non-parametric memory}
component, known as \emph{retrieval memory}. This integration is widely
referred to as Retrieval-Augmented Generation (RAG~\cite{rag-paper}), which
combines the generative capabilities of LLMs with the factual grounding
provided by retrieved documents. Naturally, the concerns mentioned
above~\cite{explainable-conversational-ir} pertaining to conventional
conversation IR systems also apply to LLM/RAG-driven systems.

In this section, we review explanations of RAG systems from two
perspectives. The first concerns understanding \emph{which parts} of the
context documents\footnote{In this section, we use the terms \emph{context
    documents} and \emph{retrieved documents} interchangeably.} contribute
to the generated \emph{answer}, and assessing how \emph{faithful} this
answer is to the retrieved evidence. Specifically, we aim to quantify the
degree to which the RAG output is grounded in its contextual documents. The
second perspective focuses on the \emph{memory} component of the large
language model, examining whether the generated answer originates from the
retrieved context or from the model's internal memory. In
Table~\ref{tab:summary-table-rag}, we summarise the approaches covered in
this section, providing a very brief description, and pointers to datasets
and codebases, where applicable.


\subsection{Faithfulness and Attributions in RAG}
In this section, we first examine the faithfulness of the generated answer --- how strongly the response is grounded in the retrieved context. We then discuss the process of attributing the output to the retrieved evidence, commonly referred to as attribution in RAG.
An answer is considered faithful if it is supported by the contextual documents; otherwise, it is deemed unfaithful. As discussed by~\citet{eavl-tacl-adlakha-etal-2024-evaluating}, faithfulness can be quantified using two main approaches: lexical-based and model-based.
Lexical-based approaches aim to assess the degree to which a RAG-generated answer / response is grounded in the retrieved knowledge with the help of statistical measures. For lexical-based approaches, they 
proposed two such measures: Knowledge-F$_1$ and Knowledge-Precision. Knowledge-F$_1$ captures the F$_1$-overlap between the answer tokens and the retrieved contexts, while Knowledge-Precision focuses on their precision alignment. However, a potential limitation of Knowledge-F$_1$ is that it may underestimate faithfulness when the generated response is concise yet accurate—since shorter answers tend to exhibit lower token overlap with lengthy context passages.
Despite their interpretability, lexical-based metrics are often surpassed by model-based metrics, which uses LLMs to estimate semantic alignment between the generated answer and the supporting documents. In the following, we discuss two such model-based evaluation frameworks proposed by~\citet{ragas-es-etal-2024} and~\citet{ares-saad-falcon-etal-2024-ares}.


In~\citet{ragas-es-etal-2024} (RAGAs), the authors extract individual statements from the generated answers and evaluate the extent to which these statements can be verified using evidence from the retrieved context documents. Both the statement extraction and verification steps are performed in a zero-shot manner using LLM. However, as the entire pipeline depends exclusively on prompt-based operations, it is inherently sensitive to prompt design and may not generalise well across different domains or tasks, thereby limiting its practical applicability. To support their evaluation, the authors constructed a new dataset comprising content from 50 Wikipedia pages. Questions were generated using ChatGPT, and human annotators provided the corresponding reference answers. The prompts used for question generation included specific instructions, such as ensuring that each question is fully supported by the given context.


\citet{ares-saad-falcon-etal-2024-ares} proposed an alternative evaluation framework, termed ARES, which leverages in-domain examples to enhance evaluation efficacy. Given an in-domain document collection, ARES first generates synthetic queries along with their corresponding answers. These synthetic pairs are then used to train LLM judges that assess whether a generated answer is faithful to its supporting context. Subsequently, the framework employs prediction-powered inference~\cite{angelopoulos2023prediction} to derive statistically grounded confidence intervals for the scores assigned by the LLM judges. Two different large language models—FLAN-T5 XXL\footnote{\url{https://huggingface.co/google/flan-t5-xxl}}
 and DeBERTa-v3-Large\footnote{\url{https://huggingface.co/microsoft/deberta-v3-large}}—were
used for synthetic data generation and for preparing the LLM judges, respectively. Experiments were conducted across a diverse range of RAG question-answering datasets, including NaturalQA~\cite{natural-qa-etal-2019-natural}, HotpotQA~\cite{yang-etal-2018-hotpotqa}, WoW~\cite{wow-dataset}, FEVER~\cite{thorne-etal-2018-fever}, MultiRC~\cite{multirc}, and ReCoRD~\cite{record}. Empirical results indicate that ARES attains a 25.9\% improvement in accuracy over RAGAs, demonstrating enhanced robustness and adaptability across domains.

\subsubsection*{Attribution} 
At a high level, the attribution problem in RAG aims to \emph{ground the generated answer} in the supporting contextual documents. One straightforward approach~\cite{bohnet2023attributedquestionansweringevaluation} formulates this as an \emph{entailment task}, where the model is shown the query, answer, and retrieved context, and is then asked whether the answer is supported by the context. However, although this formulation appears simple in principle, it fails to account for cases where the answer depends on information distributed across multiple retrieved documents, thereby limiting its applicability to more complex attribution setup. To address this issue, researchers have explored \emph{self-citation} mechanisms, where LLMs are prompted to produce citations pointing to the source documents that support their generated answers. 
Work by~\citet{attrscore-yue2023automatic} formalised these notions of citation-based attribution and categorised attributions into three distinct types:  
\begin{enumerate}[label=(\roman*)]
    \item \textbf{Attributable}: the context document directly supports the answer;  
    \item \textbf{Extrapolatory}: the context document is related but does not directly support the answer; and  
    \item \textbf{Contradictory}: the context document conflicts with the answer.  
\end{enumerate}
However, true attribution should reflect the \emph{actual reasoning process} that leads to a generated response, not merely a post-hoc citation. A citation is considered \emph{faithful} if it originates from the retrieved document set; conversely, an LLM may cite a relevant document even when the answer was produced from its internal knowledge. This phenomenon~\citet{wallat-ictir-25}, often referred to as being \emph{right for the wrong reason}, raises concerns about the reliability of citation-based attribution. 
The initial analysis by~\citet{wallat-ictir-25} showed that Cohere's RAG model\footnote{\url{https://cohere.com/blog/command-r-plus-microsoft-azure}}, optimised on the NaturalQA~\cite{natural-qa-etal-2019-natural} dataset, exhibited this behaviour in nearly 50\% of the examined cases. In their setup, documents were first retrieved using BM25 (top 30) and then re-ranked using ColBERT, with the top-5 documents subsequently fed into the RAG model. Nevertheless, the study remains constrained by its small-scale empirical evaluation. A more comprehensive analysis across multiple, well-established benchmark collections is required to validate the generality of this phenomenon.




We now discuss a fine-grained attribution framework, MIRAGE, proposed by~\citet{attribution-emnlp-qi-etal-2024-model}. 
The central idea is to identify the most important tokens in the retrieved contextual documents that have a substantial impact on the RAG output. 
This analysis is analogous to \emph{feature attribution} in the RAG pipeline, where the features correspond to tokens within the retrieved contexts. 
The method first identifies \emph{context-sensitive tokens} in the generated response -- those whose removal causes a shift in the model’s output distribution—using KL divergence. 
It then locates context words that promote these sensitive tokens through gradient-based feature attribution, selecting the top-$k$ most influential terms. A key limitation of this approach is its reliance on access to model parameters, which prevents its use with closed-source LLMs. 
Empirical findings indicate that MIRAGE achieves stronger human agreement across multiple languages on the XOR-AttriQ dataset~\cite{muller-etal-2023-evaluating}, and it outperforms self-citation methods for longer answer spans. However, MIRAGE performs less effectively for shorter answers and does not provide direct comparisons with models explicitly trained for self-citation, leaving open questions about relative efficacy across attribution paradigms. Our experiments with their framework suggest that, although MIRAGE attributes importance to contextual tokens, these attributions are often not intuitive from an end-user perspective and frequently involve tokens that appear uninformative. Nevertheless, such insights may still be valuable for model developers seeking to understand the underlying attribution process, even if they offer limited interpretability for end-users.

\subsection{Knowledge Conflicts in RAG}
In the pre-training phase, the LLMs embed diverse facts
and relationships within their parameters, forming what is referred to as the model's \emph{parametric memory}. Even in presence of contextual documents, LLM may recall few facts from knowledge and tries to answer from that. This phenomenon is denoted as the conflicts between LLM's parametric memory and the contextual documents. 
\citet{xu-etal-2024-knowledge-conflicts} introduced several key terminologies to describe different types of conflicts that can arise in such settings:  
\begin{enumerate}[label=(\roman*)]
    \item \textbf{Context–memory conflict}: a conflict between the parametric memory and the retrieved context documents;  
    \item \textbf{Inter-context conflict}: a conflict between multiple retrieved context documents; and  
    \item \textbf{Intra-memory conflict}: a case where the LLM generates inconsistent responses to semantically similar inputs.  
\end{enumerate}
These conflicts can arise due to factors such as temporal misalignment, data bias, semantic discrepancies, etc., between different knowledge sources.
Work by~\citet{conflict-bank-nips-24} proposed a large-scale benchmark for evaluating such conflicts, comprising 553K question–answer pairs and 7M claim–evidence pairs. The benchmark is constructed using Wikipedia data, where similar entity types are substituted to create three categories of conflicts: misinformation, temporal, and semantic conflicts. Experiments conducted on several LLM families—Gemma, Llama 2,  Llama 3, and Qwen 1.5—reveal that LLMs tend to prioritise external knowledge when available and exhibit notable sensitivity to conflicting evidence.


Next, we highlight a mechanism designed to resolve knowledge conflicts. 
\citet{knowledge-conflict-acl-yuan-etal-2024-discerning} proposed a two-step approach to address this issue. 
In the first step, the model identifies tokens that may conflict with the knowledge encoded in its parameters. To achieve this, it employs the entropy-based framework introduced by~\citet{arora2023stableentropyhypothesisentropyaware} to measure token prediction uncertainty by observing changes in entropy when contextual information is introduced. 
The authors observed that conflicting tokens exhibit significant entropy shifts—either markedly high or low—relative to non-conflicting tokens. Intuitively, a high entropy in token generation indicates that the model is uncertain about what to produce next.  
In the second step, an adaptive decoding mechanism is applied. During decoding, this mechanism reinforces context-aware knowledge for ``conflicting'' tokens, while relying more heavily on parametric knowledge for ``non-conflicting'' tokens. 
The process is guided by computing the divergence between token probability distributions generated with and without contextual documents. 
Experimental results on NaturalQA, SQuAD~\cite{rajpurkar-etal-2016-squad}, and StrategyQA~\cite{strategy-qa} demonstrate consistent performance across both conflicting and non-conflicting cases, indicating that the proposed mechanism can effectively reconciles discrepancies between parametric and contextual knowledge.

\begin{table}[ht]
  \centering
  \resizebox{1\textwidth}{!}{
  \begin{tabular}{l L{0.28\textwidth} L{0.25\textwidth} L{0.25\textwidth}}
    \toprule
    Approach &  Explanation type &  Dataset(s) & Codebase  \\ 
    \toprule
    RAGAs~\cite{ragas-es-etal-2024} & Quantifying faithfulness  & Created dataset & \url{https://github.com/explodinggradients/ragas} \\
    ARES~\cite{ares-saad-falcon-etal-2024-ares} & Quantifying faithfulness with confidence interval & NaturalQA, HotpotQA, WoW, FEVER, MultiRC,
ReCoRD & \url{https://github.com/stanford-futuredata/ARES} \\
    \citet{wallat-ictir-25} & Faithful attribution & NaturalQA & \url{https://github.com/jwallat/RAG-attributions} \\
    MIRAGE~\cite{attribution-emnlp-qi-etal-2024-model} & RAG attribution mechanism &  XOR-AttriQA, ELI5 & \url{https://github.com/Betswish/MIRAGE} \\
    \citet{xu-etal-2024-knowledge-conflicts} & Defined conflict categories &  Not applicable & Not applicable\\
    \citet{conflict-bank-nips-24} & Proposed conflict benchmark &  Created dataset & Not applicable \\
    \citet{knowledge-conflict-acl-yuan-etal-2024-discerning} & Resolving knowledg conflicts &  NaturalQA, SQuAD, StrategyQA & \url{https://github.com/Stacy027/COIECD} \\
    
    \bottomrule
  \end{tabular}
  }
  \caption{
  Summary table of the explanations for the RAG component discussed in Section~\ref{sec:explain-rag}.}
  \label{tab:summary-table-rag}
\end{table}


\section{Future Research}
\label{sec:future}
In the preceding sections, we have identified several issues and research
questions that have not been adequately addressed so far. We conclude in
this section by listing some of these open questions.

First, it is important to investigate how the nature of NLP tasks differs
from that of more traditional IR tasks. Traditional IR datasets (e.g.,
Robust04, ClueWeb09b) and QA-oriented datasets (e.g., MSMARCO) are clearly
different with respect to document sizes, query nature (keyword vs.\ QA),
and the presence / absence of $\langle Q, D_{+}, D_{-} \rangle$ triplets (where $D_{+}$
and $D_{-}$ are positive and negative relevant documents for a query $Q$).
These differences, combined with the fact that the Transformer models at
the core of NRMs are trained for NLP tasks, raise questions about whether
the notable success of NRMs is confined to QA-oriented datasets.
Indeed, several pre-BERT NRMs have been shown to struggle on
datasets such as Robust04 and LETOR4.0~\cite{Pang2017ADI}. In addition, BERT-based and LLM based
models have not been adequately tested on large Web collections, e.g.,
ClueWeb09b and Gov2, because of their scale.
One reason behind the limitation of BERT-based retrieval models may be
tokenization, as standard tokenizers may not be suitable for all corpora.
For example, TREC-DL 2019's medical queries necessitate adaptive
tokenizers, as explored by \citet{balde-etal-2024-adaptive} using adaptive
byte-pair encoding. In general, a more comprehensive failure analysis of
BERT-based models is needed. On a related note, domain-specific
explainability in areas like medical and legal retrieval presents unique
challenges requiring further research~\cite{sigir2022Legal}.

As discussed in Section~\ref{sec:i-or-p}, \citet{pmlr-v119-moshkovitz20a}
highlighted the difficulty of explaining cluster assignments in
high-dimensional spaces, proposing visualization using axis-aligned
features and shallow threshold trees. Applying this to explain NRMs and LLM-based rankings
 using decision tree-based explainers is a promising direction.
Given that NRMs are optimized by minimizing a loss function
over $(Q, D_{+}, D_{-})$ triplets, it is worth investigating
whether we can quantify the influence of specific triplets on the loss
function. This could reduce training costs, improve explanations,
and result in effective mining of negative documents. In fact, it is widely acknowledged in the literature that the inclusion of negative samples improves the efficacy of ranking models, and researchers have proposed various strategies~\cite{contriever, yates-etal-2021-pretrained} to extract \emph{hard negatives}. 
However, a fine-grained analysis of \emph{how} negative samples contribute to the improvement of ranking efficacy from an \emph{explainability} perspective is still lacking.


Explainability within the RAG domain represents a relatively new line of research, and significant contributions are still limited. To date, most studies have focused primarily on attribution within RAG systems and on analysing the memory component—predominantly from the perspective of LLMs. However, there has been limited investigation into how variations in document ranking influence the content or factuality of the generated responses.

One notable hypothesis in this context is the ``Lost in the Middle'' phenomenon~\cite{liu-etal-2024-lost}, which suggests that LLMs tend to under-utilise information located in the middle of the retrieved context. In particular, LLMs appear to focus more on relevant information positioned at the beginning and end of the context. This behaviour fundamentally runs counter to how standard IR metrics evaluate ranked lists. \citet{coelho-etal-2024-dwell} similarly observed that models disproportionately attend to the initial portions of the context in the MSMARCO dataset. Nevertheless, a systematic and fine-grained analysis of this effect within ranking-oriented datasets remains lacking. For example, it is still unclear whether large document collections require the top five retrieved documents to achieve high-quality responses, whereas smaller collections may suffice with only one or two. Insights of this nature could be highly valuable for practitioners seeking to optimise retrieval depth and context selection in RAG pipelines.

Another emerging direction concerns LLM-based evaluation, where LLMs act as judges to assess the quality or faithfulness of generated responses (as discussed in Section~\ref{sec:explain-rag}). While this paradigm has shown promise, it still operates largely as a ``black box'', and the principle of being right for the right reasons remains difficult to establish. Similarly, questions of faithfulness --- that is, whether the model’s attributions genuinely correspond to the supporting evidence --- have not yet been thoroughly examined. An initial effort in this direction was made by~\citet{wallat-ictir-25}, but further extensive research is needed. Building specialised datasets to support such studies would also be crucial for advancing explainability in RAG systems.

At a more basic level, the interpretability of IR evaluation measures also
warrants investigation. While some research addresses ranking constraints
for metric
axiomatics\footnote{\url{https://www.eecis.udel.edu/~hfang/AX.html}}, the
interpretability of metrics such as Average Precision and NDCG remains
under-explored, especially given NDCG's non-linear nature. 

Arguably the most pressing need of the hour, however, is a widely accepted
evaluation framework for Explainable IR / NLP, consisting of standardised
experimental protocols and test collections (see
Section~\ref{sec:eval-expl}). This would enable reliable and reproducible
comparisons of different explanation methods and foster further advances in
the field. Given the increasing emphasis on understanding complex models,
and emerging regulatory requirements, we hope that standard IR
evaluation forums (e.g., TREC, CLEF and NTCIR) will work towards
fulfilling this requirement for different categories of explanation
methods.

\nocite{volske2021towards}


\section*{Acknowledgments}
We thank anonymous reviewers for their helpful comments.  Sourav Saha is supported by TCS Research Scholar program (Cycle 17).

\nocite{trec-2019-overview}
\nocite{trec-2020-overview}

\bibliographystyle{abbrvnat}
\bibliography{survey}

\newpage
\appendix
\section{Supplementary Tables (electronic version only)}


\begin{table}[h]
  \centering
  \resizebox{0.95\textwidth}{!}{
    \begin{tabular}{l l c c}
      \toprule
      Dataset &  Descriptions &  Task & Statistics  \\ 
      \hline 
      SST & Movie Reviews & Sentiment analysis &  6,920 / 872 / 1,821\\
      IMDB & Movie Reviews & Sentiment analysis & 25,000 / 4,356\\
      WikiMovies &  Movie Domain & Question Answering & 100,000 \\
      Yelp Review Polarity & Business Reviews & Sentiment analysis & 560,000 / 38,000 \\
      M.RC & Reading Comprehension & Question Answering & 24,029 / 3,214 / 4,848\\
      SEMEVAL & Tweets & Classification & 6,000 / 2,000 / 20,630\\
      Ev.Inf & Biomedical Articles & Inference & 5,789 / 684 / 720\\
      AG & News & Classification & 102,000 / 18,000 / 7,600\\
      QQP & Quora Questions & Paraphrase & 364,000 / 391,000 \\
      QNLI & Wikipedia & QA / Natural Language Inference & 105,000 / 5,400 \\		
      MRPC & News & Paraphrase &   3,700 / 1,700  \\		
      ADR Tweets & Twitter Adverse Drug Reaction & Classification & 16,385 / 4,123\\		
      20 Newsgroups & News  & Classification & 1,426 / 334\\		
      AG News & News & Classification & 60,000 / 3,800\\		
      Diabetes (MIMIC) & Clinical Health & Classification & 7,734 / 1,614\\		
      Anemia (MIMIC) & Clinical Health & Classification & 5,098 / 1,262\\		
      CNN & News & Question Answering & 380,298 / 3,198\\		
      bAbI (Task 1 / 2 / 3)& Supporting Facts & Question Answering & 10,000 / 1,000\\		
      SNLI & Human-written Sentence Pairs & Natural Language Inference & 549,367 / 9,824 \\		
      \hline
    \end{tabular}
  }
  \caption{Details of the datasets mentioned in this survey for
    explanations of sequence models (Section~\ref{sec:sequence-models}) and
    attention (Section~\ref{sec:attention}).}
  \label{tab:dataset}
\end{table}


\section{Understanding Embeddings}
\label{sec:embeddings}
Document ranking methods optimise relevance by relying on text representations that capture semantic meaning, with word embeddings playing a crucial role by providing dense, low-dimensional vectors encoding linguistic and contextual relationships.
Word embeddings gained widespread acceptance after pre-trained embeddings,
e.g., Word2Vec~\cite{word2vec} and GloVe~\cite{pennington-etal-2014-glove},
became available. Their use as input to complex neural models is now
ubiquitous in NLP / IR. Understanding embeddings is thus key to explaining
NLP and IR methods, as they underpin how models interpret and process text.
While \textit{post-hoc} explanations attempt to explain how embeddings
capture semantics, a straightforward interpretation of individual embedding
dimensions remains an open problem. \citet{10.1145/3529755} provide a
detailed review of research on explaining word embeddings. 
This section supplements \citep{10.1145/3529755} by covering recent related
work. See Table~\ref{tab:summary-table-emb} for a summary of the articles
discussed here.

\subsection{Transformation of embedding spaces}
\label{sec:transform-emb-spaces}
A general approach to interpreting embeddings involves applying a
transformation to every word to project it into a space with interpretable
dimensions. 
The survey mentioned above \citep[Section 5.2]{10.1145/3529755} describes some of these
techniques. Below, we summarise
Densifier~\cite{rothe-etal-2016-ultradense}, a variant of which is
described in \cite{10.1145/3529755}, 
and discuss a recent study~\cite{10.1145/3366423.3380227} that was not
covered by earlier surveys.

Let $E \in \mathbb{R}^{n \times d}$ denote the embedding matrix, i.e., the stack of $n$
embedding vectors of size $d$, where $n$ is the number of words in the
vocabulary. The objective of Densifier~\cite{rothe-etal-2016-ultradense} is
to look for an orthogonal matrix $Q$ such that the product $E Q$ is
``interpretable" in the sense that the values of its first $k$ dimensions
have a good correlation with linguistic features. More precisely, suppose
$v$ and $w$ are two words and let $l$ be a linguistic feature. Intuitively,
the objective function of Densifier looks for a unit vector $q_{(l)}$ such that
$\Vert q_{(l)}^T(v-w) \Vert$ is high (resp., low) when $v$ and $w$ differ (resp., agree)
with regard to $l$. In a later work, \citet{dufter-schutze-2019-analytical}
proposed a small modification to the objective function that allows them to
obtain a closed-form analytic solution for $Q$ that is hyper-parameter
free. The transformed embeddings work well for lexicon induction and word
analogy tasks. They used a simple linear SVM and Densifier as
interpretability baselines. Both \cite{rothe-etal-2016-ultradense} and
\cite{dufter-schutze-2019-analytical} used the SemEval2015 Task 10B and
Czech movie reviews dataset; additionally,
\citet{dufter-schutze-2019-analytical} used the Word Analogy dataset.
\citet{10.1145/3366423.3380227} proposed POLAR, a framework for ensuring
the interpretability of pre-trained embedding vectors. They take a set of
$m$ polar opposite antonym
pairs. 
Suppose $(w_i,w'_i)$ is one such antonym pair, and let $v_i$, $v'_i$ be
their corresponding pre-trained embeddings. A new basis $\mathcal{B}$ is constructed
from the vectors $v_i-v'_i$ for the $m$ pairs. If $x_{d}$ is the original
pre-trained embedding of a word, then a standard transformation can be used
to compute its projection to the subspace spanned by $\mathcal{B}$.
%
POLAR~\cite{10.1145/3366423.3380227} was used to transform and evaluate 
Word2Vec~\cite{word2vec} and GloVe~\cite{pennington-etal-2014-glove} based
pre-trained word embeddings on a wide range of datasets:
News Classification, Noun Phrase Bracketing, Question Classification,
Capturing Discriminative Attributes, Word Analogy, Sentiment
Classification, and Word Similarity tasks. Across most of the tasks, POLAR
makes the vectors interpretable while retaining performance.
Interpretability was measured by comparing the top 5 dimensions of words
with Word2Vec and human annotators.

\subsection{Changing the objective function of word embeddings}
\label{change-objective-funct}
Modifying the embedding objective function is another approach to enhance interpretability. This involves incorporating components that preserve the underlying semantic structure while aligning embedding dimensions with predefined concepts. \citet{senel_utlu_sahinuc_ozaktas_koc_2021} modified GloVe's objective function to embed concepts within vector dimensions, using external concept word resources. Their cost function weighted embedding dimensions based on concept group membership, encouraging alignment with the corresponding semantic space. The embedding dimension size matched the number of concept word groups (300 from Roget's Thesaurus). However, this approach did not utilize negative vector directions. 
In follow-up work, \citet{SENEL2022102925} incorporated both positive and negative vector directions by modifying GloVe and Word2Vec loss functions. The modified cost function included components for both positive and negative concepts, assigning higher / negative weights based on positive / negative concept group membership (e.g., ``good'' and ``bad'' as opposites). Evaluations included Word Similarity and Analogy tests, along with sentiment analysis, question classification, and news classification, using SVMs trained on average input text embeddings.

\subsection{The power of contextual embedding}
\label{power-of-context-emb}
Pre-trained embeddings, such as Word2Vec and GloVe, offer context-independent word representations, which is a limitation for polysems. For example, ``apple'' has the same representation in both sentences: ``Apple has great health benefits'' and ``Apple has released a new iPad''. Context-dependent embeddings, obtained by processing entire text through models, such as CoVe~\cite{cove}, ELMo~\cite{elmo}, and BERT~\cite{devlin-etal-2019-bert}, address this issue.

\citet{tenney_iclr} investigated the advantage of contextualized embeddings by probing spans of contextual embedding vectors, concatenating them and passing them through a multi-layered network. Using CoVe, ELMo, GPT~\cite{gpt}, and BERT, they probed for sub-sentence structure. They used CoVe's GloVe vectors and ELMo's initial character-CNN activations as context-independent baselines. Experiments across various NLP tasks (part-of-speech tagging, constituent labeling, dependency parsing, named entity recognition, semantic role labeling, coreference resolution, semantic proto-role labeling, and relation classification) showed that contextualized embeddings capture syntax and long-range linguistic information better than non-contextualized ones, as measured by micro-averaged F1 scores. A randomized version of ELMo (with weights being replaced by random orthonormal matrices) performed reasonably well, but the full ELMo achieved nearly 70\% improvement, indicating ELMo's architecture plays a significant role, though its precise nature remains unexplained.

\begin{table}[ht]
  \centering
  \begin{tabular}{l L{0.3\textwidth} L{0.35\textwidth}}
    \toprule
    Approach &  Explanation type  & Codebase  \\ 
    \toprule
    Densifier~\cite{rothe-etal-2016-ultradense} & \multirow{2}{*}{Transforming word embeddings} & No URL provided \\
    \citet{dufter-schutze-2019-analytical}  &   & \url{github.com/pdufter/densray}  \\
    POLAR~\cite{10.1145/3366423.3380227} 	& Transforming word embeddings to interpretable antonyms & \url{github.com/Sandipan99/POLAR}   \\ 	
    \citet{senel_utlu_sahinuc_ozaktas_koc_2021} & Concept (positive direction) based embeddings &  \url{github.com/koc-lab/imparting-interpretability} \\
    \citet{SENEL2022102925} & Concept (positive and negative directions) based embeddings & No URL provided \\
    \citet{tenney_iclr} & Contextual vs non-contextual embeddings over their architecture & \url{github.com/nyu-mll/jiant} \\	
    \bottomrule
  \end{tabular}
  \caption{Summary of explanations for embedding vectors (Section~\ref{sec:embeddings}).} 
  \label{tab:summary-table-emb}
\end{table}


\section{Interpreting Sequence Models}
\label{sec:sequence-models}

In this section, we discuss the interpretability of sequence models, which are 
neural architectures that take input and generate output in a sequential order, maintaining representations that capture dependencies across positions. Classical sequence models include RNNs, LSTMs~\cite{lstm}, and GRUs, while modern Transformer-based architectures model sequences through self-attention rather than recurrence.
We focus on specific \emph{post-hoc} explanation approaches for LSTM  and hierarchical explanations. 
For a comprehensive discussion, we refer the reader to \citet{10.1145/3529755}.

\subsection{Predictive power of individual words}
Text classification can often be explained through key words, e.g., for intuitive tasks such as sentiment analysis, the word ``delighted'' usually indicates positive sentiment. \citet{DBLP:conf/iclr/MurdochS17} analyze LSTM outputs by measuring the influence of individual words.
Let the activation after processing the $t$-th word be defined as $h_t = o_t \odot  \mathrm{tanh}(c_t)$, where $c_t$ is the memory state, $o_t$ is the output gate and $\odot$ is the element-wise multiplication operator. After processing the full sequence of length $T$, the classification probability of the $i$-th class (among $C$ classes) is given by the softmax $p_i = \mathrm{exp}(W_{i}h_T)\big/\sum_{j=1}^C \mathrm{exp}(W_{j} h_{T})$, 
where $W$ is an appropriate parameter matrix and $W_i$ denotes the $i$-th row of $W$. \citet{DBLP:conf/iclr/MurdochS17} define the contribution of the $j$-th word towards the $i$-th class using the incremental difference of activation vectors passed through the output gate, as $\beta_{i,j} = \mathrm{exp}\left((W_{i} (o_{T}\odot (\mathrm{tanh}(c_{j}) - \mathrm{tanh}(c_{j-1}))) \right)$. Subsequently, the output of an LSTM network can be decomposed as 
\begin{align}
	\mathrm{exp}(W_{i}h_{T}) = \mathrm{exp}\left(\sum_{j=1}^{T}W_{i} (o_{T}\odot (\mathrm{tanh}(c_{j}) - \mathrm{tanh}(c_{j-1})))\right) = \prod_{j=1}^{T} \beta_{i,j}  \label{op-decomposition}
\end{align}

In a similar fashion, they define $\gamma_{i,j}$, in terms of the forget gates to reflect the upstream changes made to $c_j$ after processing the $j$-th word. 
The phrases are scored and ranked using above $\beta_{i,j}$ and $\gamma_{i,j}$ values by averaging over all the documents. The  predictive score (applying $\beta_{i,j}$ and $\gamma_{i,j}$ individually) $\mathrm{ps}_i$ 
for a class $i$ of a phrase is calculated as the ratio of average contribution of that phrase to the prediction of class $i$ to class $j$ across all occurrences of that phrase.

For experimental evaluation (binary classification), phrases were tagged with the label of the class with the maximum predictive score ($\mathrm{ps}$). 
A rule-based classifier, approximating LSTM output, performs string-based pattern matching. Given a document and a list of phrases sorted by $\mathrm{ps}$, it searches for each phrase, returning its label upon a match and stopping after the first match.

Experiments used the SST and Yelp datasets (sentiment analysis) and the Wiki Movies dataset (question answering)~\cite{DBLP:conf/iclr/MurdochS17}. For question answering, a question is encoded with an LSTM, augmented with document word embeddings, and fed to another LSTM with a softmax classifier to predict answer entities. Quantitative evaluations show the rule-based classifier can explain LSTM output via key phrases, though approximation errors exist. For example, while the LSTM correctly classified a sentence like ``Still, it \underline{gets the job done} -- a sleepy afternoon rental'' as negative, the pattern matcher tagged ``gets the job done'' as positive, suggesting the LSTM uses broader contextual information.

In a later work, \citet{james2018beyond} proposed contextual decomposition to address limitations of their prior work~\cite{DBLP:conf/iclr/MurdochS17}, which failed to identify phrases like ``used to be'' as strongly negative within a sentence. The method decomposes LSTM output and cells into phrase-specific ($\beta$) and other-factor ($\gamma$) contributions, such that the activation and memory states are represented as $h_t = \beta_t + \gamma_t$ and $c_t = \beta_t^{c} + \gamma_t^{c}$ respectively. Recursive computation groups terms derived from the phrase and external factors, identifying sentiment interactions. Contextual decomposition identifies negations and captures LSTM prediction dynamics. For example, it correctly identifies ``used to be'' as strongly negative and ``my favorite'' as positive in ``used to be my favorite,'' shown via per-word $\beta$ heatmaps. This approach was compared to cell decomposition~\cite{DBLP:conf/iclr/MurdochS17}, integrated gradients~\cite{pmlr-v119-sundararajan20b}, leave-one-out~\cite{DBLP:journals/corr/LiMJ16a}, and gradient times input through experiments conducted on Yelp and SST (binary sentiment analysis), by investigating phrase-level and word-wise CD scores.

\subsection{Feature interactions in a hierarchical structure}
Some follow-up studies aimed to identify and hierarchically use groups of the most predictive features (words), focusing on adjacent words. \citet{singh2019hierarchical} extended contextual decomposition (CD) to generic deep networks, generalizing \citet{james2018beyond}'s cell / output decomposition to layer-wise contextual decomposition: $$g^{CD}(x) = g^{CD}_{L}(g^{CD}_{L-1}(\ldots(g^{CD}_{2}(g^{CD}_{1}(x)))))$$ 
where, $x$ is the input to the network and $L$ is the number of layers in the neural network. On each layer $l$, $g^{CD}_l(x)$ is composed of two components: $\beta_{l}(x)$, measuring the importance of the feature group present in the input $x$, and $\gamma_{l}(x)$, which captures the contributions of rest of the tokens
in the input. For each layer, $g^{CD}_l(x) = \beta_{l}(x) + \gamma_{l}(x)$. One can measure the score of $\beta_{l}(x)$ and $\gamma_{l}(x)$ values on each layer (from the discussion in the earlier section~\cite{james2018beyond}).

The proposed method generalizes to CNNs, with contextual decompositions for convolution, max-pooling, dropout, and ReLU. Agglomerative contextual decomposition (ACD) was proposed to combine word groups, resembling agglomerative clustering. It combines contextual decomposition scores of each feature, merging adjacent features bottom-up via a priority queue containing words, phrases (or pixels), and their contribution scores. ACD, using SST, MNIST, and ImageNet, identifies incorrect predictions from phrase interactions. ACD is robust to adversarial perturbations (small noise added to inputs), producing similar hierarchical explanations for perturbed and original instances, capturing essential input parts.

\begin{figure}[t]
	\centering
	\begin{subfigure}[b]{0.49\textwidth}
		\centering
		\includegraphics[width=\textwidth]{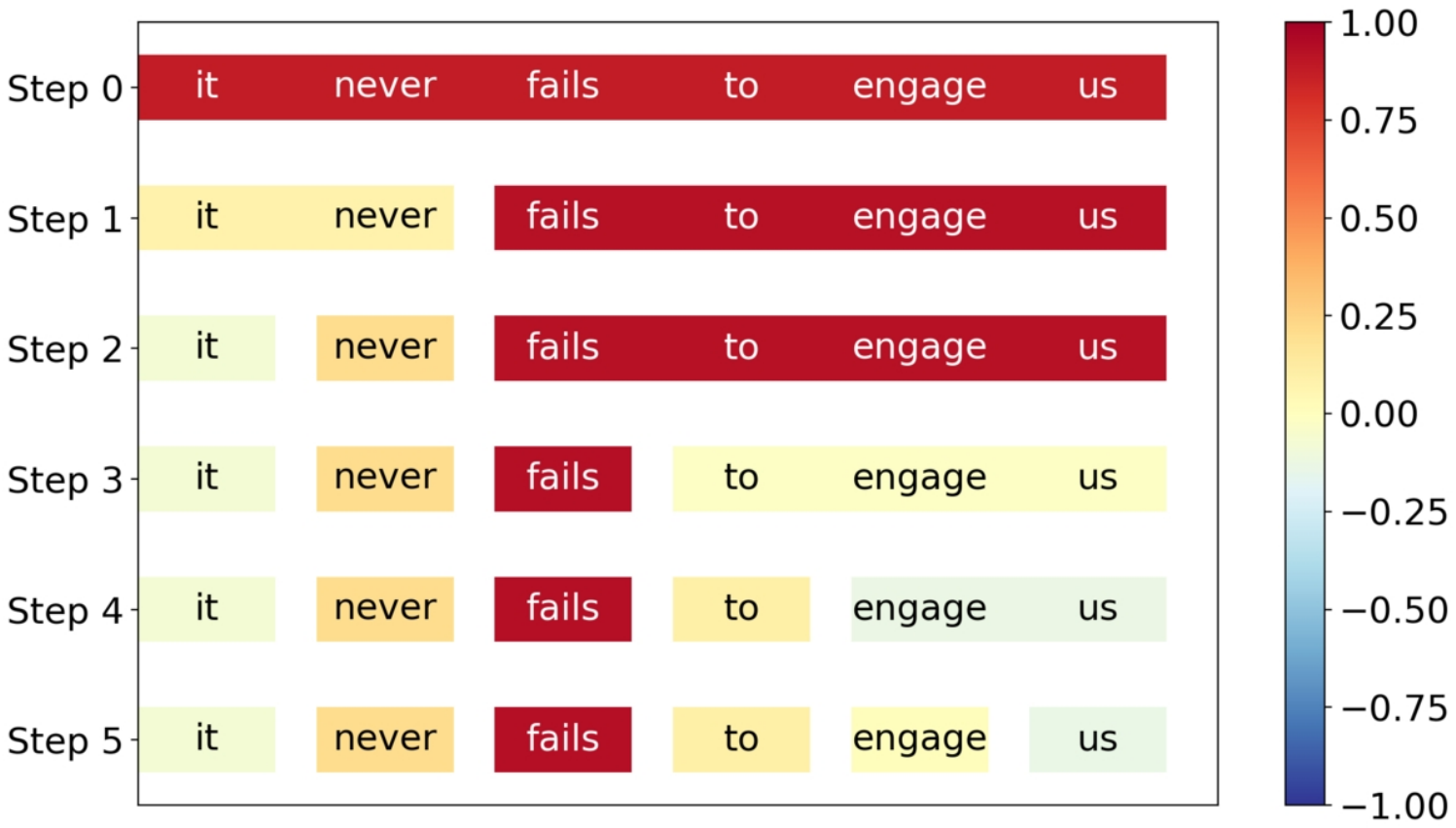}
		\caption{HEDGE method used by~\citet{chen2020generating}.}
	\end{subfigure}
	\hfill
	\begin{subfigure}[b]{0.49\textwidth}
		\centering
		\includegraphics[width=\textwidth]{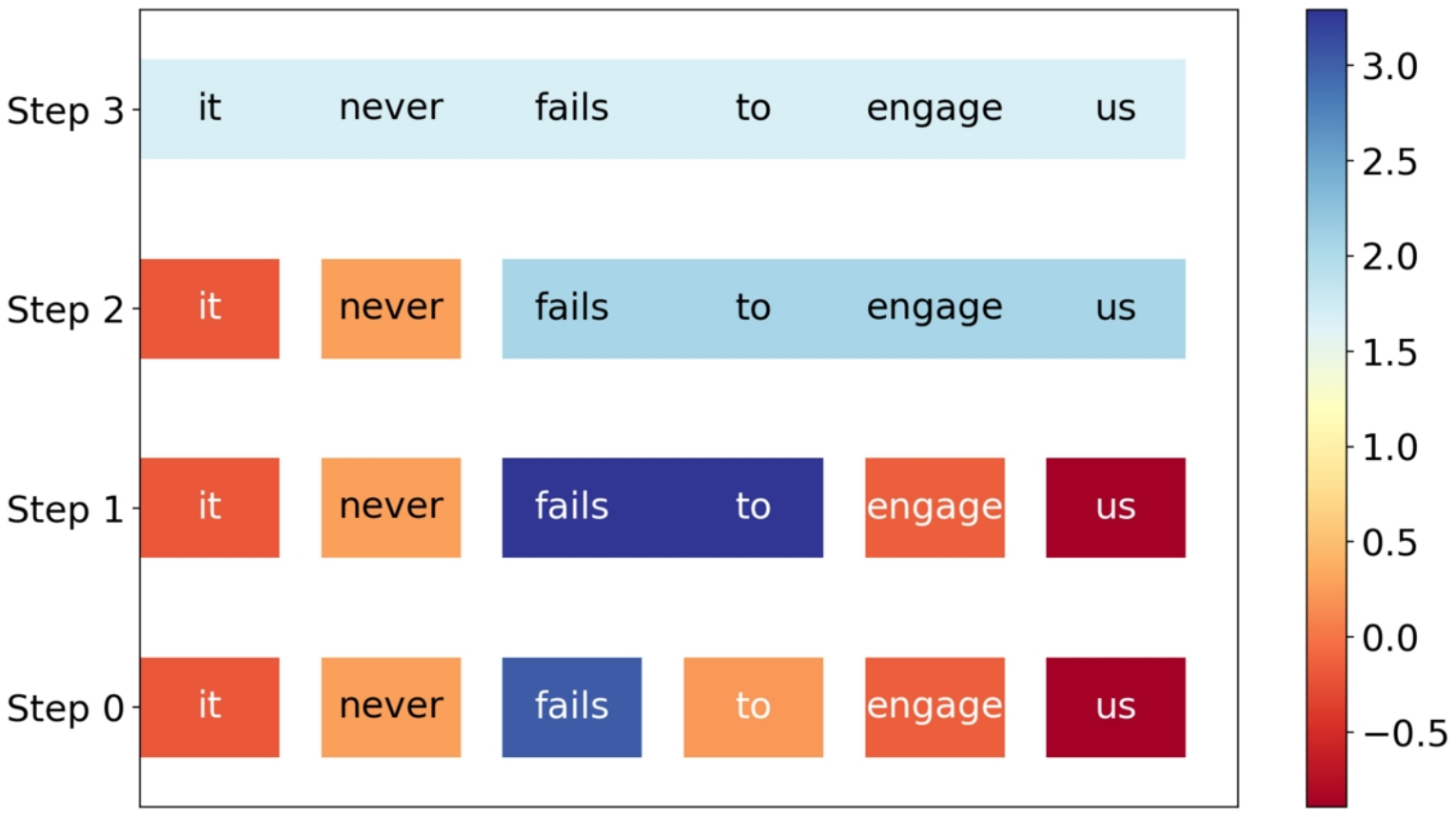}
		\caption{ACD method used by~\citet{singh2019hierarchical}.}
	\end{subfigure}
	\caption{Reproducing from ~\cite{chen2020generating}, showing the difference in interpretability results between HEDGE and ACD~\cite{singh2019hierarchical}. LSTM makes a wrong prediction as it missed the interaction between \emph{never} and \emph{fails}.}
	\label{fig:hier-interpret}
\end{figure}

HEDGE~\cite{chen2020generating} offers another hierarchical explanation. Investigating LSTM, CNN, and BERT on SST and IMDB, HEDGE iteratively divides text chunks into two segments, involving two optimizations: segment selection (outer) and cut point determination (inner). The cut point uses the Shapley interaction index~\cite{Lundberg2018ConsistentIF}. The outer optimization enumerates segments to select the one with the smallest interaction index. A segment $w_{(s_{i}, s_{i+1})}$'s contribution ($s_{i}, s_{i+1}$ are the start and end positions) measures the prediction's distance from the prediction boundary (label $y$'s contribution). Figure~\ref{fig:hier-interpret} (from~\cite{chen2020generating}) contrasts HEDGE and ACD. HEDGE initially splits sentences into larger chunks (e.g., ``it never'' and ``fails to engage us''), while ACD merges words (e.g., ``fails'' and ``to''). In this example, HEDGE’s explanations show that the LSTM makes an incorrect prediction by missing the ``never'' and ``fails'' interaction, which ACD fails to capture.

We summarise the approaches discussed in this section, along with the different datasets, and their respective codebase links, in Table~\ref{tab:summary-table-seq-model}. In Table~\ref{tab:dataset}, the relevant datasets are listed with the specific task, type of description, and dataset statistics.

\begin{table}[ht]
	\centering
    \resizebox{1\textwidth}{!}{
	\begin{tabular}{l L{0.22\textwidth} L{0.2\textwidth} L{0.25\textwidth}}
		\toprule
		Approach &  Explanation type &  Dataset(s) & Codebase  \\ 
		\toprule
		
		\citet{DBLP:conf/iclr/MurdochS17} & word-level interaction of LSTM &  WikiMovies, Yelp, SST & No URL provided \\ 
		\citet{james2018beyond}  & phrase-level interaction of LSTM & SST, Yelp  & \url{github.com/jamie-murdoch/ContextualDecomposition} \\
		ACD~\cite{singh2019hierarchical}  & Hierarchical interaction bottom up & SST  & \url{github.com/csinva/hierarchical-dnn-interpretations}  \\
		\citet{chen2020generating}  & Hierarchical interaction top down  & SST, IMDB &  \url{github.com/UVa-NLP/HEDGE} \\
				
		\bottomrule
	\end{tabular}
    }
	\caption{Summary table of the explanations for sequence models described in Section~\ref{sec:sequence-models}.} 
	\label{tab:summary-table-seq-model}
\end{table}


\section{Interpreting Attention Weights}
\label{sec:attention}
The \emph{attention mechanism}, originally termed as an \emph{alignment model} in the context of machine translation by 
~\citet{bahdanau-attention-paper}, has been a key component in text processing using deep neural networks since it was proposed. The attention models used in the literature are of two types:  additive attention~\cite{bahdanau-attention-paper} and scalar dot product~\cite{transformer}. 
We follow a similar attention model framework and notations used in~\cite{jain-wallace-2019-attention, mohankumar-etal-2020-towards} which is summarised below.

Suppose, we have a sequence of tokens $\{\myellipsis{w}{1}{T}\}$. The model inputs $x$ $\in$ $\mathbb{R}^{T\times |V|}$, containing one-hot vectors at each position of the words and $V$ is the vocabulary size. This sequence is passed over an embedding layer to get a $d$ dimensional representations of each tokens $x_e$ $\in$ $\mathbb{R}^{T\times d}$. Encoder module takes the representations and produces $T$ --$m$ dimensional hidden states. Note that one can consider LSTM, RNN or any sequence modeling network as a choice for the encoder module. Formally, the encoded representation is $h = Enc(x_e)$ $\in$ $\mathbb{R}^{T\times m}$. In a similar way we can encode a query $Q$ $\in$ $\mathbb{R}^{ m}$ (e.g., hidden representation of a question in question answering task). A similarity function $\phi$ takes $h$, $Q$ and produces scalar scores. Attention scores can be computed as: $\hat{\alpha}$ = softmax($\phi$($h, Q$)), where $\hat{\alpha}$ $\in$ $\mathbb{R}^{T}$. Context vectors are obtained by combining attention scores with the hidden states. There are mainly two common similarity functions i.e., the choice of $\phi$, i) additive $\phi$($h, Q$) = $v^{T} tanh(W_{1}h + W_{2}Q)$ and ii) scaled dot product $\phi$($h, Q$) = $\frac{hQ}{\sqrt{m}}$, where $W_{1}$, $W_{2}$ are learnable parameters and $tanh$ is the activation function.

Attention distributions can be considered as inherently explainable if there is a cause-effect relationship between higher attention and the model prediction. A higher weight implies a greater impact on the model's prediction. In general, attention based explanations are characterised as \emph{faithful} and \emph{plausible}~\cite{mohankumar-etal-2020-towards}. 
If higher attention signifies a greater impact on the model prediction, it is called as \emph{faithful} explanation.
To evaluate a \emph{faithful} explanation, we need to understand the correlation between inputs and outputs. One popular evaluation technique is to order the words based on the attention weights and compute the average number of tokens required to flip the decision of a prediction~\cite{nguyen-2018-comparing}.
On the other hand, the explanation is considered as \emph{plausible} if its weights can be understood by human. We need humans to evaluate whether they are \emph{plausible} explanation or not. 

\subsection{Debate on whether attention is explanation or not}
\label{debate-on-attention}
While attention was initially considered inherently explainable, \citet{jain-wallace-2019-attention} argued against this, conducting experiments across various datasets (SST, IMDB, ADR Tweets, 20 Newsgroups, AG News, Diabetes, Anemia, CNN News, bAbI, and SNLI) using additive attention on LSTM based models. They found 
that the correlation between attentions and feature attribution scores (gradient based measure) are weak for BiLSTMs with the attention components. Furthermore, they demonstrated that adversarial attention distributions (permutations of original attention weights) yielded similar results, suggesting attention weights are not reliable as explanations.
In a complementary study, \citet{serrano-smith-2019-attention} further examined the interpretability of attention weights. They analysed model behaviour by setting the highest attention weight to zero and observed that the overall prediction often remained unchanged. Nonetheless, they acknowledged that tokens with the highest attention weights generally exert stronger influence on model outputs compared with those with the lowest weights. Moreover, they showed that ranking tokens solely by their attention scores is suboptimal for identifying the minimal set of tokens whose removal would flip the model’s decision. Empirical results on multi-class text classification datasets such as Yelp and IMDB indicated that a considerable number of tokens need to be removed before a decision change occurs, reinforcing the conclusion that attention weights do not provide faithful explanations of model behaviour.


Further, \citet{wiegreffe-pinter-2019-attention} showed that attention weights frozen to a uniform distribution performed comparably to original weights on some datasets, suggesting attention is not essential for these tasks. 
They showed that replacing the LSTM with a multi-layered perceptron (MLP) while preserving learned attention weights resulted in significantly better performance than just a trained MLP, indicating that attention captures word-level architecture. An attempt to train an adversarial network with divergent attention distributions proved difficult and task-dependent. All experiments used binary classification tasks with LSTMs on datasets including SST, IMDB, 20 Newsgroups, AG News, Diabetes, and Anemia.

\subsection{Impact of sparse attention on interpretability}

\label{sparse-attention}
Recent work has explored sparse attention~\cite{pmlr-v48-martins16, malaviya-etal-2018-sparse} to reduce computational cost, with the idea that attending to fewer tokens could also improve interpretability. However, \citet{meister-etal-2021-sparse} showed that sparseness does not guarantee better interpretability. Using the Sparsegen~\cite{sparsegen} projection function to create sparse attention distributions (controlled by sparsity parameters), they investigated the relationship between input tokens and their co-indexed intermediate representations - more specifically, how much each input token influences the magnitude of these representations. They also checked if these representations could be primarily linked to a single input part, proposing a normalized entropy of gradient-based feature importance scores. A low entropy (e.g., 0) indicates a unique influence. Experiments on IMDB, SST, and 20News (topic classification) using LSTM and Transformer models suggested that the relationship between intermediate representations and input tokens is not one-to-one, even with sparse attention. They found that adding sparsity decreases plausibility and that sparse attention correlates less with other feature importance measures compared to standard attention.

\subsection{Faithful and plausible attention distributions by changing the loss function}
\label{change-loss-function-attention}
A body of work has attempted to foster more faithful and plausible attention distributions. One approach incorporates faithfulness by modifying the model's objective / loss function with a penalty term, applicable to attention distributions and sequence model hidden states. \citet{mohankumar-etal-2020-towards} proposed Orthogonal and Diversity LSTMs, observing that vanilla LSTMs' hidden states often exhibit similarity, occupying a narrow cone in latent space and yielding similar predictions even with permuted attention weights. Orthogonal LSTM orthogonalizes hidden states using the Gram–Schmidt process, while Diversity LSTM adds a conicity measure as a reward factor to the loss function, penalizing similar hidden state vectors.  The conicity measure for a set of vectors $V = \{\myellipsis{v}{1}{n}\}$ is defined as: 
\[ \mathit{conicity}(V) = \frac{1}{n} \sum_{i=1}^{n} \cos(v_i, \frac{1}{n} \sum_{j=1}^{n} v_j). \]

A high \textit{conicity} value indicates that the hidden state vectors are highly similar, resulting in low diversity among the representations. Such high conicity often explains why LSTM representations are not fully faithful to the model’s predictions. Consequently, the permutation of attention weights~(Section~\ref{debate-on-attention}) tends to yield similar contextual vectors.
Evaluating these modified LSTMs on twelve datasets (text classification, NLI, paragraph detection, and QA) showed that Orthogonal and Diversity LSTMs had much lower conicity than vanilla LSTMs. Diversity LSTM's extracted rationales received more focused attention and were shorter. This model is more human-understandable and correlates well with integrated gradients~\cite{10.5555/3305890.3306024}, reducing attention on punctuation and improving distribution across parts of speech. Diversity LSTMs focus more on nouns and adjectives, which are important for tasks like sentiment analysis.

\citet{chrysostomou-aletras-2021-enjoy} modified the Transformer loss function to incorporate faithfulness. Using TEXTRANK~\cite{mihalcea-tarau-2004-textrank} to compute word salience distributions, they penalized the model when attention deviated from these distributions using KL divergence. Experiments on SST, AGNews, EV.INF, M.RC, and Semeval showed improved faithfulness compared to rationale extraction and input erasure.

Table~\ref{tab:summary-table-attention} provides summary, datasets, and the codebase of the explanation approaches discussed in this section. The details of the datasets related to the attention based explanations can be found in Table~\ref{tab:dataset}.

\begin{table}[ht]
	\centering
    \resizebox{1\textwidth}{!}{
	\begin{tabular}{l L{0.2\textwidth} L{0.3\textwidth} L{0.25\textwidth}}
		\toprule
		Approach &  Explanation type &  Datasets & Codebase  \\ 
		\toprule		
		\hline
		\citet{jain-wallace-2019-attention} & Claims attention is not explainable  & SST, IMDB, ADR
Tweets, 20 Newsgroups, AG News, Diabetes, Anemia, CNN, bAbI, SNLI & \url{github.com/successar/AttentionExplanation}  \\
        \citet{serrano-smith-2019-attention} & Claims attention is not explainable  & Yahoo Answers, IMDB, Amazon, Yelp & \url{github.com/serrano-s/attn-tests}  \\

		\citet{wiegreffe-pinter-2019-attention} & Argues attention maybe explainable  &  SST, IMDB,
20 Newsgroups, AG News, Diabetes, Anemia &  \url{github.com/sarahwie/attention} \\
		\citet{meister-etal-2021-sparse} & Sparsify attention / attend to fewer tokens & IMDB, SST, 20 Newsgroups & No URL provided  \\
		\citet{mohankumar-etal-2020-towards} & Faithful and plausible attention with modified LSTM & SST, IMDB, Yelp, Amazon, Anemia, Diabetes, 20 Newsgroups, Tweets, SNLI, QQP, babI, CNN  & \url{github.com/akashkm99/Interpretable-Attention} \\
		\citet{chrysostomou-aletras-2021-enjoy} & Faithful transformer based predictions & SST, AG News, EV.INF, MultiRC & \url{github.com/GChrysostomou/saloss} \\
		
		\hline
		
		\bottomrule
	\end{tabular}
    }
	\caption{Summary table of the explanations for the attention module discussed in Section~\ref{sec:attention}.}
	\label{tab:summary-table-attention}
\end{table}


\section{Interpreting Transformers and BERT}
\label{sec:transformers-bert}
Since the invention \todo{invention: too strong a word?} of the
Transformer~\cite{transformer} and subsequently
BERT~\cite{devlin-etal-2019-bert}, the landscape of neural NLP and IR has
changed significantly, as BERT based models have been very effective on
various NLP and IR tasks. Consequently, a number of researchers have
focused on the interpretability of different components of Transformer and
BERT. The survey by \citet{rogers-etal-2020-primer} provides an aggregated
view of the research done in this topic.
Broadly, research questions regarding the explainability of BERT can be
categorised into the following classes:
\begin{enumerate*}[~(1)]
    \item The significance of different layers of BERT,
    \item Investigation on BERT's capability (or inability) to learn linguistic knowledge, 
    \item The role of different attention heads, and 
    \item Analyses of the information flow in BERT.
\end{enumerate*}
In this section, we provide an overview of the research efforts categorized
into these aspects. Additionally, Table~\ref{tab:bert-findings} summarises
the findings reported in these articles.

\begin{table}[ht]
    \centering
    \begin{tabular}{p{7.5cm} p{6.5cm}}
    \toprule
    \textbf{Findings} & \textbf{Citations (Codebases)} \\
    \otoprule
    BERT-base beats BERT-large on subject-verb agreement and sentence subject detection. & \citet{goldberg2019assessing} (\url{github.com/yoavg/bert-syntax}) and  ~\citet{lin-etal-2019-open} (\url{github.com/yongjie-lin/bert-opensesame})\\ \hline 
    BERT struggles to adjust over negation. & \citet{ettinger-2020-bert} (\url{github.com/aetting/lm-diagnostics}) \\ \hline 
    BERT does not comprehend arguments. & \citet{niven-kao-2019-probing} (\url{github.com/IKMLab/arct2}) \\ \hline 
    Most of the effectiveness of Transformer and BERT are due to a few of the attention heads, as pruning the rest does not significantly reduce the performance of the models. & \citet{voita-etal-2019-analyzing} (\url{github.com/lena-voita/the-story-of-heads}) and \citet{Budhraja2021OnTP} (NA) \\ \hline 
    Attention heads in the earlier layers are local, i.e., tokens attend to nearby tokens. 
    There are some evidences that the attention heads perform different roles to learn different linguistic knowledge.     
    & \citet{clark2019what} (\url{github.com/clarkkev/attention-analysis}), \citet{DBLP:conf/aaai/PandeBNKK21} (\url{github.com/iitmnlp/heads-hypothesis}) \\ \hline 

    The middle layers of BERT are multi-skilled heads, later layers are specific to task. & \citet{DBLP:conf/aaai/PandeBNKK21} \\ 
    \bottomrule
    \end{tabular}
    \caption{A summary of what BERT can do and why, as reported in the
      papers discussed in Section~\ref{sec:transformers-bert}.} 
    \label{tab:bert-findings}
\end{table}

\subsection{Significance of different layers of BERT}
\label{sig-diff-layers-bert}
As expected, larger BERT variants (with more layers and hidden states) generally outperform base variants~\cite{devlin-etal-2019-bert}, though exceptions exist, such as subject-verb agreement~\cite{goldberg2019assessing} and sentence subject detection~\cite{lin-etal-2019-open}, where the base model performs better. These anomalies are rare, and it is unclear if the larger model is inherently less potent or if further fine-tuning could improve its performance. 

Several studies investigate what aspects different layers of deep architectures encode, typically using layer-specific word representations in probing tasks to determine captured syntactic representations. \citet{blevins-etal-2018-deep} demonstrated that RNNs encode some syntax without explicit supervision, using probing classifiers on different RNN layers to predict parts-of-speech and constituent label ancestry (parent, grandparent, great-grandparent). Using English Universal Dependencies, CoNLL-2012, WMT02014 English-German, and CoNLL-2012 datasets for dependency parsing, semantic role labeling, machine translation, and language modeling, they trained deep RNNs and observed that higher layers encode less syntactic structure and more abstract representations, based on layer-wise prediction accuracy on various syntactic tasks.

In a spirit similar to \cite{blevins-etal-2018-deep}, \citet{tenney-etal-2019-bert} proposed two scoring functions to evaluate BERT layers. 
The first function, namely scalar mixing weight, is designed to evaluate the importance of individual layers.
 With $l$ layers and $j^{th}$ layer representation $y_{j}$, the final decision is based on $\sum \alpha_{j} y_{j}$, where $\alpha_{j}$ is tuned on training data to determine the importance of layers. 
 Another scoring formula, namely cumulative score, denotes how much the accuracy measure F1 score changes for the introduction of a specific layer of BERT. Term $Score(y_{j})- Score(y_{j-1})$ can be used to measure the changes in F1 score observed while adding the $j^{th}$ layer. 
 
 Experiments on English Web Treebank, SPR1, SemEval, and OntoNotes 5.0 for eight tasks (POS, constituents, dependencies, entities, semantic role labeling, coreference, semantic proto-roles, and relation classification) showed BERT follows a similar trend to RNNs: basic syntax is captured earlier, and later layers are better for semantics. Syntactic structures are generally more localizable, while semantics are spread across layers. However, \citet{tenney-etal-2019-bert} noted cases where syntactic identification occurs in later layers. For example, in ``china today blacked out a cnn interview that was...'', initial layers tag ``today'' as a common noun (date). Due to ambiguity, the model later updates ``china today'' to a proper noun, changing its semantic role. In this case, the model performs syntactic, then semantic, and finally both syntactic and semantic tasks.

\subsection{BERT's capability (or inability) to learn linguistic knowledge}
\label{sec:bert-linguistic}
\citet{jawahar-etal-2019-bert} identified linguistic structures captured by BERT base, confirming that surface structures are encoded in initial layers, syntactic structures in middle layers, and semantic information in upper layers, using probing tasks (SentEval~\cite{conneau-kiela-2018-senteval}), as discussed in Section~\ref{sec:probing}. Experiments across ten probing tasks considered surface features (sentence length), syntactic features (tree depth), and semantic features (tense)~\cite{conneau-etal-2018-cram}. BERT also captures dependency structures, with higher layers used for long-range dependencies. \citet{tpdn-network} used a Tensor Product Decomposition Network (TPDN) to validate whether BERT implicitly represents \textit{role-filler} pairs (e.g., in sequence copying, $\{3, 8, 7, 5\}$ as $3:first + 8:second + 7:third + 5:fourth$). They used syntactic tree position (path from root) as the role and the word as the filler, assuming that if a TPDN can approximate a neural network's learned representation for a role scheme (by minimizing mean squared error), the network implicitly captures that scheme. Several researchers have investigated BERT's limitations, showing it struggles with argument comprehension~\cite{niven-kao-2019-probing} and negation~\cite{ettinger-2020-bert}.

\subsection{Understanding multi-head attention}
Vaswani et al.~\cite{transformer} claimed that multi-head attention allows attending to information from different representation subspaces, unlike single-head attention. Subsequent studies investigating Transformer and BERT attention maps have partially confirmed this. \citet{voita-etal-2019-analyzing} analyzed Transformer attention maps for machine translation, showing that only a few heads are responsible for performance, suggesting many can be pruned. \citet{Budhraja2021OnTP} adopted this for multilingual BERT. \citet{clark2019what} investigated BERT-base attention maps, probing each head for specific syntactic relations on dependency parsing datasets. They observed that earlier layers contain heads attending heavily to previous / next tokens. Attention to [SEP] is largely inconsequential, while broad attention occurs in lower layers and the last layer's [CLS] token. Attention maps represent English syntax, distributed across multiple heads. Some heads behave similarly, and heads within the same layer often have similar distributions. While probing showed some heads focus on specific dependency relations (e.g., coreference, preposition / object, verb / direct object), individual heads do not clearly attend to the overall dependency structure.

\begin{figure}[t]
	\centering
	\includegraphics[scale=0.3]{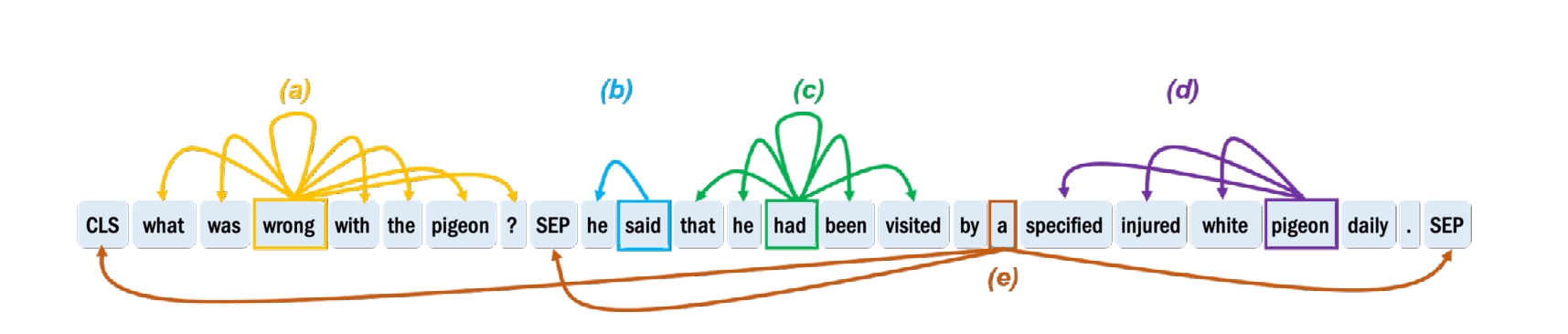}
	\caption{An example (reproduced from
          \cite{DBLP:conf/aaai/PandeBNKK21}) showing the attention sieve received by various
          input tokens. The current tokens are marked with boxes and color
          coded edges point to other tokens in the sieve for different
          functional roles: a) block, b) nsubj (syntactic), c) local, d)
          amod (syntactic), and e) delimiter.} 
	\label{fig:func-label-bert}
\end{figure}

In a follow up work, \citet{DBLP:conf/aaai/PandeBNKK21}  formalized the analysis by classifying the attention heads based on their functional roles, into the following categories:
\begin{enumerate*}[~(a)]
\item \emph{local}: tokens are attended in a small neighborhood of current token, generally next / previous tokens.  
\item \emph{syntactic}: attending tokens which are syntactically related to the current token.
\item \emph{delimiter}: attending special tokens, i.e., [CLS] and [SEP]. 
\item \emph{block}: tokens are attended within the same sentence contrary to the tokens in the sentences before or after the [SEP] token.
\end{enumerate*}
Figure~\ref{fig:func-label-bert} shows this categorization with a sample sentence.
The authors defined \emph{attention sieve} as a set of tokens that will be
attended to obtain the representation for a specific attention role. They
also defined a \emph{sieve bias score} as the ratio between the average
attention scores given to the \textit{tokens in the sieve} and to
\textit{all tokens in the input sequence}. In case of syntactic attention
heads, the set of tokens from the input sequence that make up a sieve was
identified by a dependency parser. For other heads it is quite
straightforward to determine the sieve tokens. To identify attention heads
that satisfy specific functional roles, \citeauthor{DBLP:conf/aaai/PandeBNKK21}
used thresholds on the sieve bias score. Their proposed framework provides
an option for tuning the strictness of the classification of attention
heads by adjusting the thresholds.
The higher the threshold value, the stricter the classification rule for designating the particular head.
Experiments on QNLI~\cite{wang-etal-2018-glue} (QA Natural Language
Inference), QQP~\cite{wang-etal-2018-glue} (paraphrase detection),
MRPC~\cite{mrpc} (paraphrase detection), and
SST-2~\cite{socher-etal-2013-recursive} (sentiment analysis)suggest that
middle layers of BERT have multi-skilled heads, while later layers are task-specific. Syntactic heads are also local heads and delimiter heads overlap with other functional heads.  

\subsection{Information flow in BERT}
\label{information-flow-bert}
\citet{quantifying-attention-flow} attempted to quantify attention flow in Transformers. In a later work, \citet{Lu2020InfluencePF} observed that most of the information is passing through the skip connections of BERT, and showed that attention weights alone are not sufficient enough to quantify information flow. 
Using a gradient-based feature attribution~\cite{10.5555/3305890.3306024} method that has been discussed earlier, \citet{Lu2020InfluencePF} introduced influence patterns as how gradient flows from input to output through skip connections, dense layers, etc. Gradient based influence patterns from input to output token can be constructed considering all possible paths using the chain rule. To reduce the computational overhead, they have used strategies to greedily add influential nodes from one layer to the adjacent layer. A path from input tokens to the output is considered a form of explanation.
Experimental findings were conducted on subject word agreement and reflexive anaphora tasks. While previous methods only look into the aspects of attention weights, they do not discuss how information is traveling through the skip connections.  
In this work, they argued that BERT uses attention to contextualize information and skip connections to pass that information from one layer to another. Once contextualize step is finalized it relies more upon the skip connections. Experimental results also show that BERT is able to do negation in sentiment analysis tasks very well. 

\par
Another information interaction in the Transformer network was studied by~\citet{hao2021self} through the lens of self-attention attribution~\cite{10.5555/3305890.3306024} based method. With feature attribution (integrated gradients) techniques, they identified important attention connections i.e., each pair of words in the attention heads of different layers. Intuitively, pairs of words receive a high score if they contribute more to the final prediction. It was observed that a larger attention score between a pair of words may not contribute towards the final prediction. 

\begin{figure}
	\centering
	\begin{subfigure}[b]{0.49\textwidth}
		\centering
		\includegraphics[width=\textwidth]{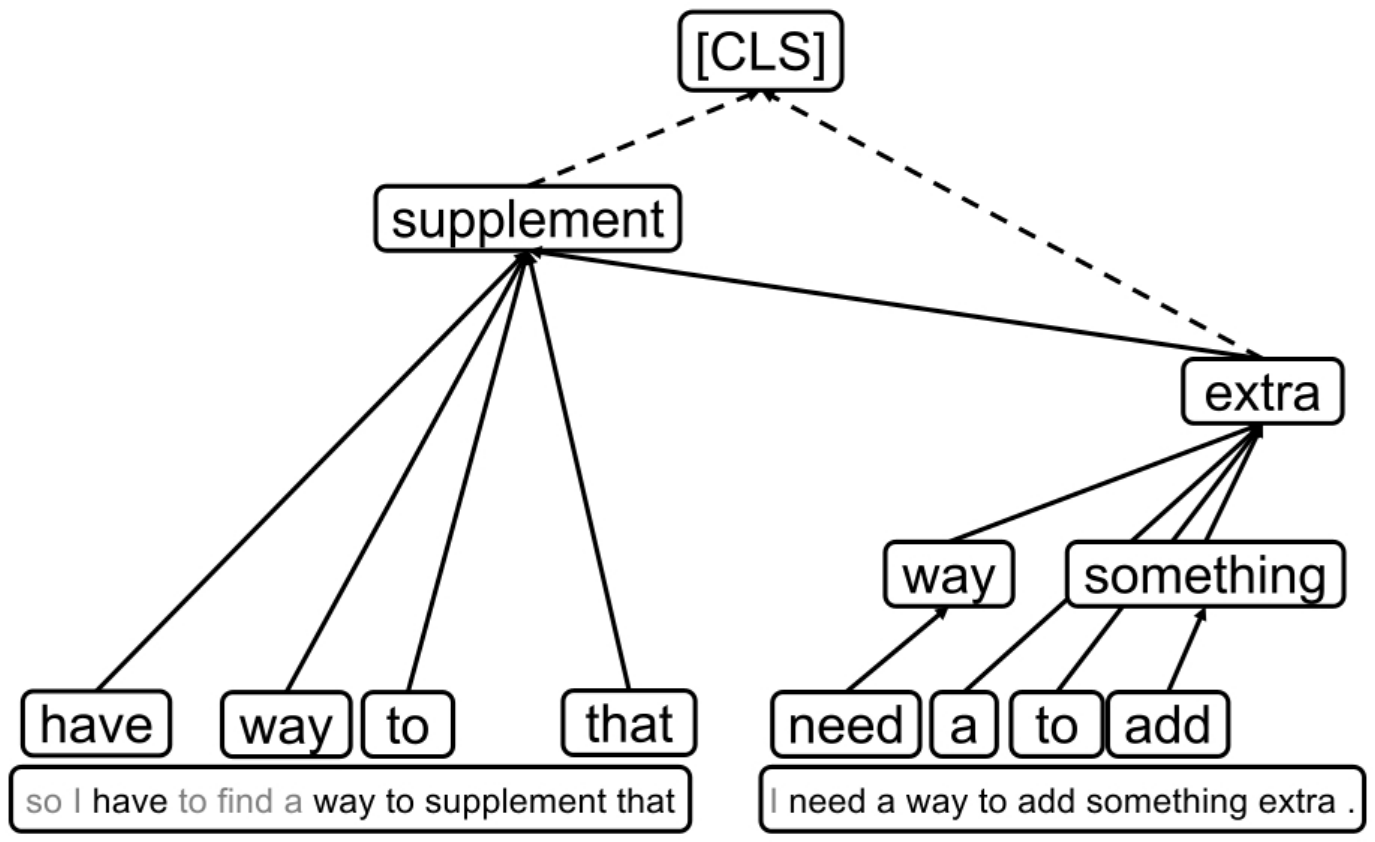}
		\caption{Attribution tree generated from MNLI dataset, where the final prediction is `entailment'.}
	\end{subfigure}
	\hfill
	\begin{subfigure}[b]{0.49\textwidth}
		\centering
		\includegraphics[width=\textwidth]{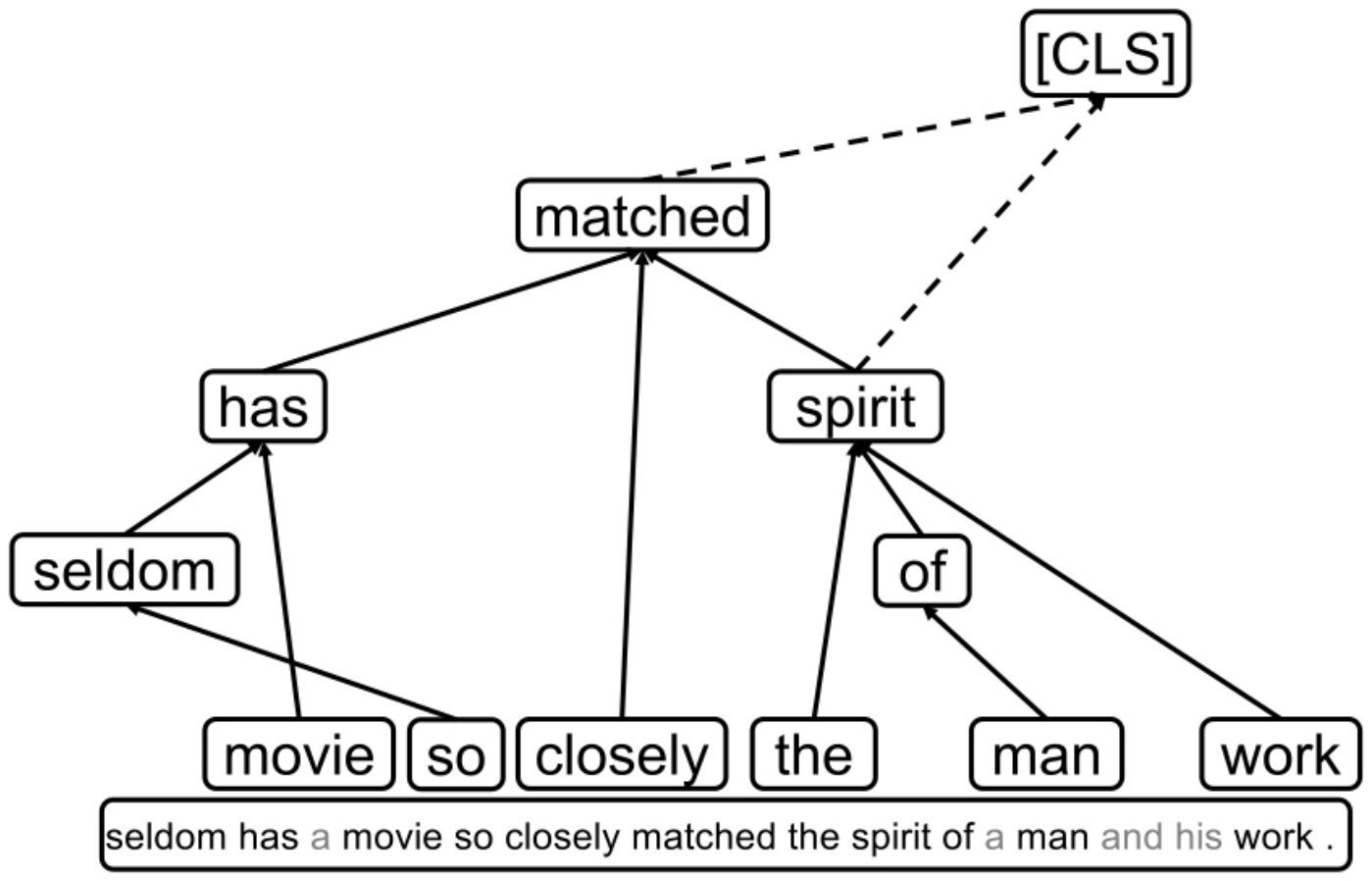}
		\caption{Attribution tree generated from SNLI-2 dataset; the prediction of the sentence is `positive'.}
	\end{subfigure}
	\caption{Reproducing figures from~\citet{hao2021self}, showing the examples of attribution trees generated with different datasets.}
	\label{fig:attribution-ex}
\end{figure}

With the attribution scores of different heads, they pruned attention heads from different layers of BERT. Accuracy values do not change much if the heads with small attribution scores were pruned; however, heads with large attribution scores significantly impact the performance. In general, it was observed that two heads across each layer produce a good performance, and pruning the heads with average attention scores does not significantly affect performance. Further, important attention heads are highly correlated across different datasets. All the empirical results were conducted on MNLI, RTE, SST-2, and MRPC datasets.
A heuristic algorithm was designed to demonstrate the information flow between input tokens inside the BERT layer. Figure~\ref{fig:attribution-ex} shows two instances, one from the MNLI dataset and another from the sentiment analysis dataset SST-2. For the first one, the interactions between input sentences try to \emph{explain} the prediction. However, for the second example, although `seldom', `movie', and `man' are more important words for the `positive' sentiment to a human; the information flows presented by their approach do not aggregate to these tokens. 




%

\end{document}